\documentclass[aps,preprintnumbers,prl,twocolumn,superscriptaddress]{revtex4}
\usepackage{graphicx}         
\usepackage{bm}               
\usepackage{amssymb}          
\usepackage{amsmath}          
\usepackage{mathtools}
\usepackage{lipsum, color}

\usepackage{amssymb,amsmath,mathrsfs,epstopdf,slashed,color}
\usepackage{tikz}
\usetikzlibrary{matrix}
\usetikzlibrary{positioning}
\usepackage{fancyhdr} 
\usepackage{lastpage} 
\usepackage{extramarks} 
\usepackage{graphicx} 
\usepackage{lipsum} 
\usepackage{amsmath}
\usepackage{amsfonts}
\usepackage{amssymb}
\usepackage{latexsym}
\usepackage{color}
\usepackage{xypic}
\usepackage{cancel,slashed}

\usepackage{graphicx}
\usepackage{placeins}
\usepackage{hyperref}
\usepackage{braket}
\usepackage{bm}
\usepackage{diagbox}
\usepackage{multirow}
\usepackage{newfile}

\newcommand{\diag}{\textrm{diag}}

\allowdisplaybreaks

\graphicspath{{./figures/}}
\usepackage{xcolor}
\usepackage{xspace}

\begin{document}
\title{The Hubble Constant in the Axi-Higgs Universe}

\author{Leo WH Fung}
\email{whfungad@connect.ust.hk}
\affiliation{Department of Physics and Jockey Club Institute for Advanced Study, \\
The Hong Kong University of Science and Technology, Hong Kong S.A.R., China}

\author{Lingfeng Li}
\email{iaslfli@ust.hk}
\affiliation{Department of Physics and Jockey Club Institute for Advanced Study, \\
The Hong Kong University of Science and Technology, Hong Kong S.A.R., China}

\author{Tao Liu}
\email{taoliu@ust.hk}
\affiliation{Department of Physics and Jockey Club Institute for Advanced Study, \\
The Hong Kong University of Science and Technology, Hong Kong S.A.R., China}

\author{Hoang Nhan Luu}
\email{hnluu@connect.ust.hk}
\affiliation{Department of Physics and Jockey Club Institute for Advanced Study, \\
The Hong Kong University of Science and Technology, Hong Kong S.A.R., China}

\author{Yu-Cheng Qiu}
\email{yqiuai@connect.ust.hk}
\affiliation{Department of Physics and Jockey Club Institute for Advanced Study, \\
The Hong Kong University of Science and Technology, Hong Kong S.A.R., China}

\author{S.-H. Henry Tye}
\email{sht5@cornell.edu}
\affiliation{Department of Physics and Jockey Club Institute for Advanced Study, \\
The Hong Kong University of Science and Technology, Hong Kong S.A.R., China}
\affiliation{Department of Physics, Cornell University, Ithaca, NY 14853, USA}

\begin{abstract}

		The $\Lambda$CDM model provides an excellent fit to the CMB data. However, a statistically significant tension emerges when its determination of the Hubble constant $H_0$ is compared to the local distance-redshift measurements. The axi-Higgs model, which couples an ultralight axion to the Higgs field, offers a specific variation of the $\Lambda$CDM model. It relaxes the $H_0$ tension as well as explains the $^7$Li puzzle in Big-Bang nucleosynthesis, the clustering $S_8$ tension with the weak-lensing data, and the observed isotropic cosmic birefringence in CMB. In this paper, we demonstrate how the $H_0$ and $S_8$ tensions can be relaxed simultaneously, by correlating the axion impacts on the early and late universe. In a benchmark scenario ($m=2 \times 10^{-30}$ eV) selected for experimental tests soon, the analysis combining the CMB+BAO+WL+SN data yields $H_0 = 69.9 \pm 1.5$~km/s/Mpc and $S_8 = 0.8045 \pm 0.0096$. Combining this (excluding the SN~(supernovae) part) with the local distance-redshift measurements yields $H_0 = 72.42 \pm 0.76$~km/s/Mpc, while $S_8$ is slightly more suppressed.

\end{abstract}

\maketitle

\section{Introduction}

One of the greatest successes in cosmology is the precise measurements of cosmic microwave background (CMB) which support the inflationary-universe paradigm combined with the $\Lambda$-cold-dark-matter ($\Lambda$CDM) model. However, as data improves, a significant discrepancy emerges between its Hubble constant $H_0=67.36 \pm 0.54$ km/s/Mpc determined by Planck 2018 (P18)~\cite{Aghanim:2018eyx}, and $H_{0}=73.3 \pm 0.8$ km/s/Mpc obtained from the local distance-redshift (DR) measurements (see~\cite{Verde:2019ivm} and also~\cite{Knox:2019rjx,DiValentino:2021izs} and references therein). At the same time, the $\Lambda$CDM/P18 data fitting gives the clustering $S_8=0.832 \pm 0.013$~\cite{Aghanim:2018eyx}, while the recent weak-lensing (WL) data of KiDS-1000 and DES yield $S_8=0.766^{+0.020}_{-0.014}$~\cite{Heymans:2020gsg} and $0.773^{+0.026}_{-0.020}$~\cite{Abbott:2017wau}, respectively.

The axi-Higgs model recently proposed~\cite{Fung:2021wbz} suggests a potential solution to the $H_0$ tension by coupling ultralight axions to the Higgs field. This model can further explain the $^7$Li puzzle in Big-Bang nucleosynthesis (BBN) (by shifting the Higgs vacuum expectation value (VEV)~\cite{Kneller:2003xf, Li:2005km,Coc:2006sx,Dent:2007zu, Browder:2008em,Bedaque:2010hr,Cheoun:2011yn,Berengut:2013nh,Hall:2014dfa,Heffernan:2017hwa,Mori:2019cfo}), as well as the CMB $S_8$ tension with the WL data and the observed isotropic cosmic birefringence (ICB) in CMB~\cite{Minami:2020odp}. In this paper, we will demonstrate how this model resolves the $H_0$ and $S_8$ tensions simultaneously, by correlating axion impacts in the early and late universe. 

Keeping all parameters in the standard model of particle physics unchanged, the axi-Higgs model with a single axion-like particle $a$ and the electroweak Higgs doublet $\phi$ is given by
\begin{eqnarray}\label{model1}
	V(a, \phi, \phi^{\dagger})  &=& m^2 a^2/2 + \left|m_s^2F(a) - \kappa \phi^{\dagger} \phi\right|^2 \ ,\\
F(a) &=& (1+\delta v)^2 = 1+ C{a^2}/{M_{\rm Pl}^2} \ , 
\end{eqnarray}
where the axion mass $m$ is  $\sim 10^{-30} - 10^{-29}$~eV~\cite{Fung:2021wbz}; $m_s$ and $\kappa$ are fixed by the Higgs VEV today $v_0=246$ GeV and the Higgs mass $m_{\phi}=125$ GeV; $\delta v=\Delta v/v_0=(v-v_0)/v_0$ is the fractional shift of $v$ from $v_0$; $M_{\rm Pl}=2.4 \times 10^{18}$\,GeV is the reduced Planck mass.

The perfect square form of $V(a, \phi, \phi^{\dagger})$ here is crucial. It suppresses the impact of the Higgs evolution, such that the axion evolves as if it is free~\cite{Fung:2021wbz}. In this context, the axion (1) lifts $v$ (and hence electron mass $m_e$~\footnote{Massive particles couple to the Higgs field usually. But, only electrons are relevant here: the protons receive a contribution to their mass dominantly from strong dynamics, while massive elementary particles except electrons are too heavy to affect recombination significantly. }) in early universe (while keeping the Higgs energy density unshifted~\cite{Fung:2021wbz}) and relaxes it to today's value in late universe, (2) contributes to dark matter (DM) density in the later universe and impacts the comoving diameter distance, (3) explains the ICB data with its Chern-Simons coupling to photons, (4) with its super-long de Broglie wavelength, dampens the clustering amplitude, and (5) provides observable tests via atomic clock and quasar spectral measurements.

Concretely, the axion field stays in a misaligned initial state $a_{\text{ini}}$ until the Hubble parameter $H(z)$ drops to $\sim m$. Then $a$ rolls down along the potential, and deposits its vacuum energy into DM. This occurs at the redshift $z_a \sim 1000 - 100$, and determines today's relic abundance $\omega_a$. Since $z_\text{BBN} \sim 10^9$ at BBN while $z_* \sim 1100$ at recombination, $a_{\text{ini}}$ yields $\delta v_{\text{ini}}=\delta v_{\text{BBN}}=\delta v_{\text{rec}}$ throughout the BBN-recombination epoch. Replacing the coefficient $C$ and $a_{\text{ini}}$ by $\delta v_{\text{ini}}$ and $\omega_a$ ($C \simeq 0.12 \delta v_{\text{ini}}/\omega_a$), we have five parameters to determine: the baryon density $\omega_b$, the DM density $\omega_c$ (excluding the $\omega_a$ contribution), $h=H_0/100$~km/s/Mpc, $\delta v_\text{ini}$ and $\omega_a$, with
\begin{equation}
h^2 = \sum \omega_i = \omega_b + \omega_c + \omega_a + \omega_\Lambda \ .   \label{eq:freq}
\end{equation}

Considering that the share of $\omega_a$ is tiny and its effects on standard cosmological parameters are typically of percent level, we will apply a leading-order perturbative approach (LPA)~\cite{Fung:2021wbz} in this study. While it is well-known that establishing an (semi-)analytical relation between Hubble constant and model parameters would be important for revealing the underlying mechanism to address the Hubble tension (see, e.g.,~\cite{Jedamzik:2020zmd,Sekiguchi:2020teg}), a systematic method for achieving this goal has been missing. The LPA strongly responds this need, allowing us to clearly see how a cosmological model like ``axi-Higgs'' interplays with the observation data, with relatively small computational effort.

    \begin{table}[htp]
\resizebox{0.49\textwidth}{!}{
			\begin{tabular}{|c|cc|cc|}
				\hline
				\multirow{2}{*}{} & $\Lambda$CDM & axi-Higgs & $\Lambda$CDM & axi-Higgs \\ 
				& \multicolumn{2}{c|}{(CMB+BAO+WL+SN)} & \multicolumn{2}{c|}{(CMB+BAO+WL+DR)} \\ 
				\hline\hline
				$\omega_b$ & $0.02251 \pm 0.00013$ & $0.02272 \pm 0.00020$ & $0.02267 \pm 0.00013$ & $0.02299 \pm 0.00014$ \\
				
				$\omega_c$ & $0.11801 \pm 0.00081$ & $0.1205 \pm 0.0019$ & $0.11657 \pm 0.00077$ & $0.1228 \pm 0.0014$ \\
				
				$H_0$ & $68.24 \pm 0.36$ & $69.9 \pm 1.5$ & $68.96 \pm 0.34$ & $72.42 \pm 0.76$ \\
				
				$v_{\rm ini}/v_0$ & $1$ & $1.0123 \pm 0.0086$ & $1$ & $1.0254 \pm 0.0050$ \\
				
				$1000\omega_a$ & $0$ & $< 1.25~[95\%]$ & $0$ & $< 1.32~[95\%]$ \\
				\hline

				$\omega_\Lambda$ & $0.3244 \pm 0.0056$ & $0.345 \pm 0.020$ & $0.3356 \pm 0.0053$ & $0.377 \pm 0.010$ \\
				
				$\omega_m$ & $0.14116 \pm 0.00078$ & $0.1444 \pm 0.0022$ & $0.13988 \pm 0.00075$ & $0.1470 \pm 0.0017$ \\
				
				$S_8$ & $0.8084 \pm 0.0093$ & $0.8045 \pm 0.0096$ & $0.7902 \pm 0.0088$ & $0.7970 \pm 0.0088$ \\
				
				$100\theta_*$ & $1.04129 \pm 0.00029$ & $1.04115 \pm 0.00030$ & $1.04154 \pm 0.00028$ & $1.04107 \pm 0.00030$ \\
				
				\hline
				$N_Y - N_X$ & $19 - 3$ & $19 - 5$ & $21 - 3$ & $21 - 5$ \\
				\hline
				
			\end{tabular}
}
		\caption{Marginalized parameter values in the $\Lambda$CDM model and the axi-Higgs model with $N_Y$ is the number of observables and $N_X$ is the number of model parameters. The upper bounds of $\omega_a$ are shown at 95\% confidence level.} 
		\label{tab:model_comparison}
	\end{table}

The effects of varying the other cosmological parameters are sub-leading. So we simply fix them to the default/best-fit values of $\Lambda$CDM/P18~\cite{Aghanim:2018eyx}. These parameters include: the density $\omega_{\nu}$ for two massless neutrinos and one light one ($m_{\nu} = 0.06\,$eV), $A_s = 2.10055\times 10^{-9}$ and $n_s = 0.96605$ of the initial curvature spectrum, and the reionization optical depth $\tau_\text{re} = 0.05431$. Notably, for $m \simeq 10^{-30} - 10^{-29}~{\rm eV}$, the axion perturbations affect the low-$l$ plateau of the CMB spectra with a level below that of cosmic variance~\cite{Marsh:2015xka, Hlozek:2014lca, Hlozek:2017zzf,FrancoAbellan:2021hdb}, while their effects in the high-$l$ region which are characterized by a sub-Jeans scale are essentially suppressed. The axion perturbations thus can be safely neglected in the LPA analysis here. Note that this feature is not shared by the model of early dark energy~(EDE)~\cite{Karwal:2016vyq,Poulin:2018cxd}, where the favored axion is relatively heavy ($m \simeq 10^{-27}$~eV; see, e.g.,~\cite{Poulin:2018cxd, Lin:2019qug, Agrawal:2019lmo}) and its perturbation effects hence may not be negligible~\cite{Poulin:2018dzj} (see supplemental material at Sec.~C for details). 

We summarize the main analysis results in Tab.~\ref{tab:model_comparison} with $m=2 \times 10^{-30}$ eV as an axi-Higgs benchmark. Combining the CMB+BAO(baryon acoustic oscillation)+WL+SN(supernovae) data yields $H_0 = 69.9 \pm 1.5$~km/s/Mpc and $S_8 = 0.8045 \pm 0.0096$. Especially, $\delta v_{\rm ini}=1.23$\% agrees well with $\delta v_{\text{BBN}}=1.2$\% required to solve the $^7$Li puzzle \cite{Fung:2021wbz}. The DR data further up-shifts $H_0$ to $72.42 \pm 0.76$~km/s/Mpc with $\delta v_{\rm ini} = 2.54$\%, which is higher than needed by BBN. This tension however can be solved by introducing a second axion~\cite{Fung:2021wbz}, conveniently the one ($m \simeq 10^{-22}$ eV) for fuzzy DM~\cite{Hu:2000ke,Schive:2014dra,Hui:2016ltb}.

	\section{The LPA Analysis}
	
We separate the LPA analysis into the following steps: (1) determine the set of parameters ${\bf X}$ characterizing the relevant model (${\bf X} \equiv \{ \omega_b; \omega_c; h; v_{\rm ini}; \omega_a \}$ for the axi-Higgs model) and a collection of compressed observables ${\bf Y}$ representing the data, where $N_Y \ge N_X$; (2) define a reference point (here we choose the best fit in the $\Lambda$CDM/P18 scenario~\cite{Aghanim:2018eyx} as the reference point, $i.e.$, ${\bf X}_{\rm ref} = \{ 0.02238; 0.1201; 0.6732; v_0; 0 \}$), with the observable reference values ${\bf Y}_{\rm ref} = {\bf Y} \left( {\bf X}_{\rm ref} \right) $;  (3) derive variation equations of these observables w.r.t. ${\bf X}$ at the reference point: 
		\begin{equation}
			\delta Y \equiv d\ln Y = Y_{|b} \delta \omega_b + Y_{|c} \delta \omega_c + Y_{|h} \delta h + Y_{|v_{\rm ini}} \delta v_{\rm ini} + Y_{|a} \omega_a \ ,  \label{eq:gen}
		\end{equation} 
where $Y_{|X} \equiv \partial\ln Y / \partial\ln X$, with an exception of $Y_{|a} \equiv \partial\ln Y / \partial \omega_a$, and their values are calculated either analytically from their definition or numerically using public Boltzmann codes; and (4) apply a likelihood method to these variation equations, to find out the parameter values favored by data. 

The likelihood function is defined as 
	\begin{equation}
		\mathcal{L} = 
		\frac{1}{\sqrt{(2\pi)^{N_Y} | \mathbf{\Sigma}|}} \exp  \left[ -\dfrac{1}{2} ({\bf Y}_\text{o} - {\bf Y}_{\rm t})^\text{T} \mathbf{\Sigma}^{-1} ({\bf Y}_\text{o} - {\bf Y}_{\rm t}) \right]  \ , 
		\label{eq:lik}
	\end{equation}
with  ${\bf Y}_{\rm t}= {\bf Y}_{\text{ref}} \left( 1 + \delta {\bf Y} \right)$. Here the subscripts ``o'' and ``t'' represent the observation values and model predictions, respectively. $\mathbf{\Sigma}$ is covariance matrix, given by $\mathbf{\Sigma}_{ij} = \rho_{ij} \sigma_i\sigma_j $, where $\rho_{ij}$ measures the observable correlation with $\rho_{ij} = \rho_{ji}$ and $\rho_{ii} = 1$, and $\sigma_i$ is observation variance. The numerical MCMC sampler \texttt{Cobaya}~\cite{Torrado:2020dgo} is used to sample the likelihood function for our analysis.

	\begin{table}[htp]
\resizebox{0.35\textwidth}{!}{
		\centering
		\begin{tabular}{ |c|c|c|c|c|c| }
		\hline
		\diagbox{Y}{X}  & $ \omega_b$ & $ \omega_c$ &  $ h$ & $ v_{\rm ini}$ & $\omega_a$  \\	
		\hline 			
	    $ l_a$ & $0.0550$ & $- 0.1203$ & $- 0.1934$ &  $0.6837$ & $- 2.6364$ \\		
		\hline
		$ l_\text{eq}$ & $0.0942$ & $0.5082$ & $-0.1934$ & $0.0154$ & $- 2.6376$   \\
		\hline
		$ l_D$ & $0.2459$ & $- 0.0962$ & $- 0.1934$ & $0.4553$ & $-2.6315$ \\		
		\hline
		$S_L$ & $-0.1398$ & $0.8484$ & $-0.2619$ & $0.0788$  &  $-16.082$  \\		
		\hline
		$ \alpha_\perp$(0.698) & $0.1466$ & $0.0969$ & $-0.7175$ &  \multirow{3}{*}{$0.6162$} & $-0.9852$  \\
		\cline{1-4}\cline{6-6}
		$ \alpha_\parallel$(0.698) & $0.1256$ &$-0.0160$ & $-0.4483$ &  & $-1.9252$  \\
		\cline{1-4}\cline{6-6}
		$ \alpha_V$(0.845) & $0.1354$ &  $0.0370$ & $-0.5747$ &  & $-1.4838$  \\
		\hline
		$ S_8$ & $-0.1007$ & $1.0578$ & $-0.7658$ & $0.0788$  &  $-14.337$  \\
		\hline
		$ m_B (1.36)$ & $-0.0018$ & $-0.0099$ & $0.0237$ & $0$ &  $-0.0829$  \\
		\hline
		\end{tabular}
		}
		\caption{Representative $Y_{|X}$ values in the axi-Higgs model. The $\Lambda$CDM(+$m_e$) model shares the values of  $Y_{|b,c,h} (+Y_{|v_{\rm ini}})$. }
		\label{tab:par_derivs_final}
	\end{table}

To represent the CMB data, we consider the sound horizon at recombination $l_a$, the Hubble horizon at matter-radiation equality $l_{\rm eq}$, the damping scale at recombination $l_D$ and $S_L$. $l_a$ determines the position of the first acoustic peak and also the peak-spacings. $l_\text{eq}$ sets up the threshold for radiation to dominantly drive gravitational potential, while $l_D$ is the scale below which fluctuations are suppressed by photon-baryon coupling and multipole anisotropic stress. Phenomenologically, $l_\text{eq}$ and $l_D$ determine the relative peak heights while $l_\text{eq}$ also determines the modulation between the even and odd peaks. As pointed out in~\cite{Hu:1995kot, Hu:2000ti, Hu:2001bc}, the CMB temperature spectrum $C^{\rm{TT}}_{\ell}$ can be effectively characterized by these scale parameters. The CMB polarization spectrum $C^{\rm{EE}}_{\ell}$ and cross spectrum $C^{\rm{TE}}_{\ell}$ measure similar acoustic features~\cite{Aghanim:2018eyx} and can couple to these scales also. As for $S_L \equiv \sigma_8 \Omega_m^{0.25}$~\cite{Planck:2018lbu}, it encodes the CMB lensing spectrum $C^{\rm{\phi\phi}}_{\ell}$ and reflects the CMB constraints on matter fluctuation. Moreover, we include the BAO scale parameters in the direction transverse ($\alpha_\perp (z_{\rm eff})$) and parallel ($ \alpha_\parallel (z_{\rm eff})$) to the line of sight respectively and the isotropic BAO scale parameter ($\alpha_V (z_{\rm eff})$), the galaxy-clustering amplitude ($S_8$) from WL, and the supernova luminosity ($m_B(z_{\rm eff})$) (or the local DR measurements). Conveniently, we denote ${\bf Y_{\rm CMB}} = \{l_a, l_\text{eq}, l_D, S_L \}$, ${\bf Y_{\rm BAO}} = \{\alpha_\perp (z_{\rm eff}), \alpha_\parallel (z_{\rm eff}), \alpha_V (z_{\rm eff})\}$, ${\bf Y_{\rm WL}} = \{S_8\}$, ${\bf Y_{\rm SN}} = \{m_B(z_{\rm eff})\}$ and ${\bf Y_{\rm DR}} = \{H_0\}$. The data respectively applied to them include: 
\begin{itemize}
	\item {\bf CMB}: P18 (TT,TE,EE + lowE + lensing) data~\cite{Aghanim:2018eyx};
	\item {\bf BAO}: low-$z$ surveys 6dF~\cite{Beutler_2011}, MGS~\cite{Ross:2014qpa} and high-$z$ eBOSS ELG~\cite{Raichoor:2020vio} for $\alpha_V$; high-$z$ eBOSS LRG~\cite{Bautista:2020ahg}, Quasar~\cite{Neveux:2020voa}, Lyman-$\alpha$~\cite{duMasdesBourboux:2020pck} for $\alpha_\perp, \alpha_\parallel$;
	\item {\bf WL}: DES Y1~\cite{Abbott:2017wau} and KiDs-1000~\cite{Heymans:2020gsg};
	\item {\bf SN}: Binned Pantheon samples~\cite{Scolnic:2017caz};
	\item {\bf DR}: SH0ES-19~\cite{Riess:2019cxk}, H0LiCOW~\cite{Wong:2019kwg}, MCP~\cite{Reid:2008nm}, CCHP~\cite{Freedman:2019jwv}, SBF~\cite{Potter:2018} and MIRAS~\cite{Huang:2018dbn}.
\end{itemize}
Combining these data yields a block-diagonal covariance matrix: ${\bf \Sigma} = \diag \{{\bf \Sigma_{\rm CMB}},{\bf \Sigma_{\rm BAO}},{\bf \Sigma_{\rm WL}},{\bf \Sigma_{\rm SN/DR}}\}$.

The $Y_{|X}$ values are presented in Tab.~\ref{tab:par_derivs_final} (see supplemental material at Sec.~A  and Sec.~B for a full list of ${\bf Y}$ and ${\bf \Sigma}$, and $Y_{|X}$ respectively). The relevant variation equations then can be read out directly, using these $Y_{|X}$ values as the inputs of Eq.~(\ref{eq:gen}). For example, we have
\begin{eqnarray}\label{eq:dleq}
 \delta l_{\rm eq}  	&=& 0.0942 \delta\omega_b + 0.5082 \delta\omega_c - 0.1934\delta h \nonumber \\ 
	&& +0.0154 \delta v_{\rm ini} - 2.6376 \omega_a  
\end{eqnarray}
for $Y=l_{\rm eq}$. Since $l_{\rm eq}$ has been precisely measured, a shift in $h$ has to be compensated for by shifts in the other quantities, to keep $\delta l_{\rm eq}\simeq 0$. Separately, with $S_{8|a} \ll 0$, lowering $S_8$ needs only a small $\omega_a$.

		\begin{table}[htp]
\resizebox{0.48\textwidth}{!}{
		\centering
\begin{tabular}{ |c|c|c|c|c| }	
\hline	
	${\bf X}$	& $\omega_b$ & $\omega_c$ & $h$ & $S_8$ \\ \hline
	${\bf X}_{\rm P18}$~\cite{Aghanim:2018eyx}	& $0.02237\pm 0.00015$  & $0.1200\pm 0.0012$ & $0.6736\pm 0.0054$ & $0.832 \pm 0.013$ \\ \hline
	${\bf X}_{\rm LPA}$	& $0.02237\pm 0.00014$ & $0.1200\pm 0.0011$ & $0.6735 \pm 0.0050$ & $0.832 \pm 0.013$  \\ \hline		
\end{tabular}
		}
		\caption{Test of the LPA validity in the $\Lambda$CDM model. }
		\label{tab:CMBdata}
		\end{table}

\begin{figure} [!ht]
		\centering
		\includegraphics[scale=0.4]{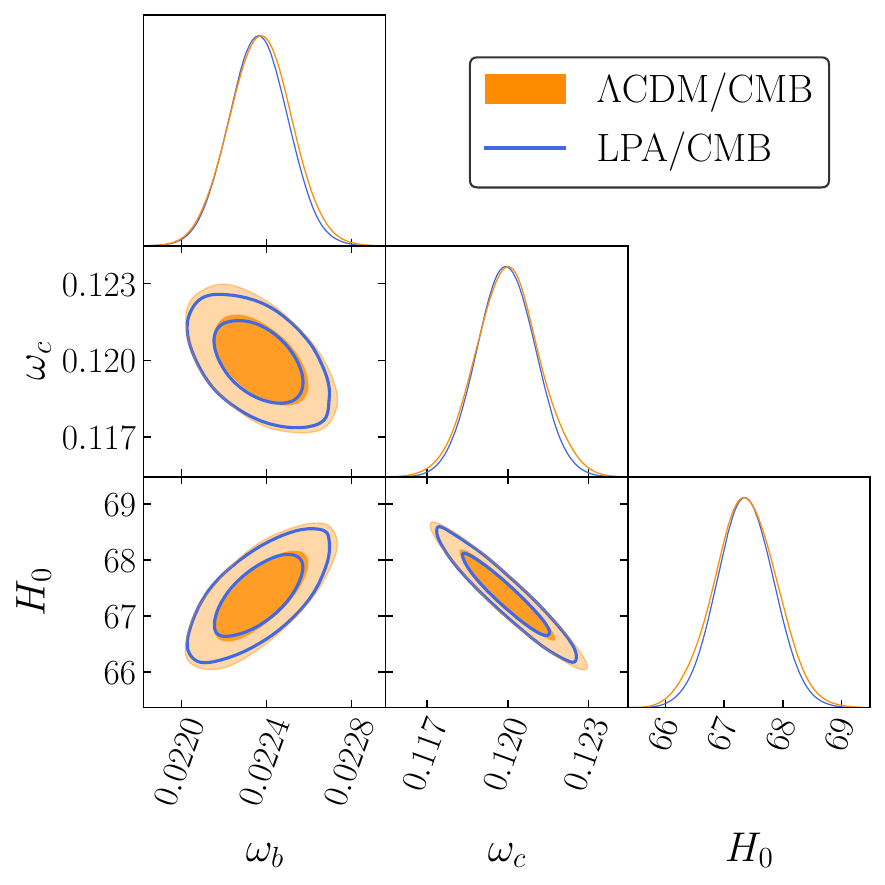}
		\caption{Comparison of the marginalized contours and posterior distributions of $\Lambda$CDM between the LPA and the P18's~\cite{Aghanim:2018eyx} analyses. }
		\label{fig:LPA}
	\end{figure}

We first test the LPA validation with ${\bf Y_{\rm CMB}}$ in the $\Lambda$CDM model. As shown in Fig.~\ref{fig:LPA}, the LPA exceptionally reproduces the marginalized contours of $\omega_b$, $\omega_c$ and $H_0$ and their posterior distributions obtained by P18~\cite{Aghanim:2018eyx}. Numerically, the LPA results (${\bf X}_{\rm LPA}$) only differ from the marginalized $\Lambda$CDM/P18 ones (${\bf X}_{\rm P18}$) slightly, for both central values and their uncertainties (see Tab.~\ref{tab:CMBdata}). The LPA is equally successful while being applied to the $\Lambda$CDM$+m_e$ model~\cite{Ade:2014zfo,Hart:2019dxi}, where $\delta v_{\rm ini} = \delta m_e$. This provides an even more crucial test on the LPA validity as this model is characterized by its own parameters. The LPA validity is then expected for the axi-Higgs model (see supplemental material at Sec.~D for details): in terms of cosmological phenomenology, the axi-Higgs model differs from $\Lambda$CDM$+m_e$ mainly in the impacts on the comoving distance to last scattering. Note, for other cosmological models, the LPA validity needs to be further tested.

\section{$\mathbf{H_0}$ and $\mathbf{S_8}$ in the Axi-Higgs Universe}

Despite the sharing of $m_e$ as a free parameter in the recombination epoch, the axi-Higgs model is essentially different from $\Lambda$CDM+$m_e$, due to the impacts of the time-varying axion field. According to~\cite{Ade:2014zfo, Hart:2017ndk}, an upward shift of $m_e$ will reduce the cross section of Thomson scattering ($\propto m_e^{-2}$) and modify various atomic processes crucial to recombination. It thus increases $z_*$ and decreases the comoving sound horizon and damping scale at $z_*$. The axion evolution here causes a positive shift to $v$ or $m_e$ at $z > z_a$ and then brings it back later to its today's value~\cite{Fung:2021wbz}. In contrast, such a mechanism is lacking for $\Lambda$CDM+$m_e$~\cite{Ade:2014zfo, Hart:2017ndk}. Moreover, the axion at $z < z_a$ tends to reduce the comoving diameter distance, because of its contribution to $H(z)$. This provides extra flexibility to resolve the impacts of varying $H_0$ on the CMB scale parameters. As to be shown below, a combination of these effects raises $H_0$ to a value higher than what the $\Lambda$CDM+$m_e$ model allows, without breaking our knowledge on the today's electron.

	\begin{figure}[!ht]
		\centering
		\includegraphics[scale=0.48]{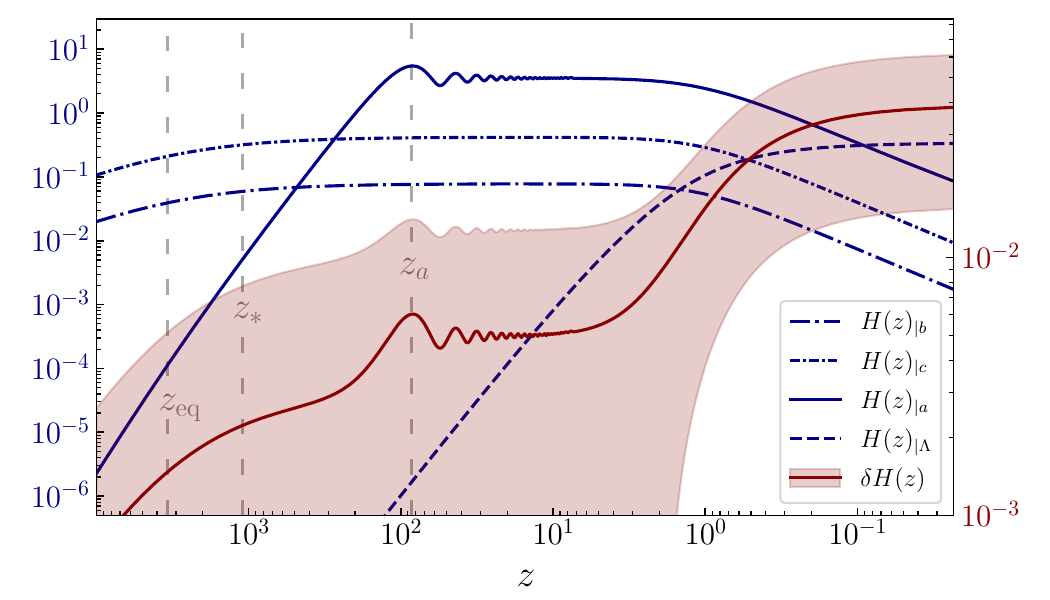}
		\caption{Evolution of $H(z)_{|b,c,a,\Lambda}$ and $\delta H(z)$ in the axi-Higgs benchmark, with the CMB+BAO+WL+SN data. $z_{\rm eq}$ denotes the redshift at the moment of matter-radiation equality.}
		\label{fig:phpw}
	\end{figure}

Let us start with the Hubble flow of axi-Higgs:
\begin{eqnarray}
&H(z) &= \ 100 \text{ km/s/Mpc} \times \\
 &&[\omega_r (1+z)^4 + (\omega_c+\omega_b) (1+z)^3 + g(z) \omega_a + \omega_{\Lambda}]^{\frac{1}{2}} \ ,  \nonumber 
\end{eqnarray} 
where $\omega_r$ is the radiation density, and 
\begin{equation}\label{eq:a1osc}
  g(z) = \begin{dcases}
    (1+z)^3 \ ,  & z \ll z_a<z_*  \ ; \\
   (1+z_a)^3 \ ,  & \quad  z > z_a  \ . 
  \end{dcases}
\end{equation}
The evolution of $\delta H(z) = (H(z) - H_{\rm ref}(z)) / H_{\rm ref}(z)$ and its derivatives w.r.t. $\omega_{b, c, a,  \Lambda}$ then can be derived from this formula. We show both of them in Fig.~\ref{fig:phpw} using the best-fit of $\Lambda$CDM/P18 as the reference scenario as before. According to this figure, $H(z)$ deviates from its $\Lambda$CDM prediction since $z>z_{\rm eq}$, which is sequentially taken over by $\omega_c$, $\omega_a$ and $\omega_\Lambda$. The evolution of $H(z)_{|a}$ can be separated into three stages. In the early time, the axion is dark energy(DE)-like. $H(z)_{|a}$ evolves as $\propto 1/ \omega_r (1+z)^4$ for $z>z_{\rm eq}$ and $\propto 1/(\omega_b +\omega_c)(1+z)^3$ after that. So its value is suppressed at high redshift. This lasts until the axion becomes DM-like at $z \sim z_a$. $H(z)_{|a}$ then evolves roughly as a constant $\propto 1/\omega_m$ during $z \sim 100 - 1$, with a wiggling feature developed for its curve due to axion oscillation. In the $\Lambda$-dominant epoch ($z < 1$), $H(z)_{|a}$ drops quickly as $z$ goes to zero, as $H(z)_{|a} \propto {(1+z)^3}/ {\omega_\Lambda}$. Such an evolution pattern of $H(z)_{|a}$, particularly its big value after recombination, results in a universal negative dependence of the CMB and BAO scale parameters on $\omega_a$ (see Tab.~\ref{tab:par_derivs_final} and supplemental material at Sec.~B for details). Consider $l_{\rm eq} \propto H(z_{\rm eq}) D_*$ as an example. $H(z_{\rm eq})$ is determined by the early-time cosmology and hence less influenced by $\omega_a$, while the diameter distance $D_* =\int_{0}^{z_*} \frac{dz'}{H(z')}$ is closely related to cosmic evolution after recombination, varied as $\sim \int_{0}^{z_*} \frac{ - H(z')_{|a} \omega_a dz'}{H(z')^2}$ w.r.t. $\omega_a$. So we necessarily have $l_{{\rm eq}|a} < 0$ (as a comparison, we have $l_{{\rm eq}|c} > 0$). Note, both $H(z_{\rm eq})$ and $D_*$ and hence $l_{\rm eq}$ are insensitive to $\delta v_{\rm ini}$.

In $\Lambda$CDM, an $H_0$ value from local DR measurements is highly disfavored by the CMB data due to its correlation with $\omega_b$ and $\omega_c$. The Friedman equation for today's universe (see Eq.~(\ref{eq:freq})) indicates that, as $h$ increases, $\omega_i$ tends to increase faster. Being out-of-phase between these variations breaks the variation equations of the CMB/BAO scale parameters defined by Tab.~\ref{tab:par_derivs_final}. However, the situation gets changed in the axi-Higgs model. Among these scale parameters, varying $h$ tends to have the largest impact on $l_{\rm eq}$ via $\delta \omega_c$. This impact is largely cancelled by the $\omega_a$ contribution. As discussed above (also see Eq.~(\ref{eq:dleq})), $\delta l_{\rm eq}$ has an opposite dependence on $\delta \omega_{c}$ and $\omega_a$. As for the impacts brought in by the requested $\omega_a$ on the other scale parameters, they will be absorbed by a positive $\delta v_{\rm ini}$ which also compensates for the impacts of shifting $h$. Except $l_{\rm eq}$, these parameters demonstrate a positive and comparable dependence on $\delta v_{\rm ini}$, due to the universal impacts of $\delta v$ on the sound horizon at recombination and the end of baryon drag. The interplay of these parameters finally mitigates the $H_0$ tension. Notably, although the effect of varying $m_e$  in the $\Lambda$CDM+$m_e$ model can be encoded as that of $\delta v_{\rm ini}$ here, the absence of $\omega_a$ worsens the fitting of $l_{\rm eq}$ and hence limits the allowed values for $h$ greatly.

		\begin{figure} [!ht]
		\centering
		\includegraphics[scale=0.28]{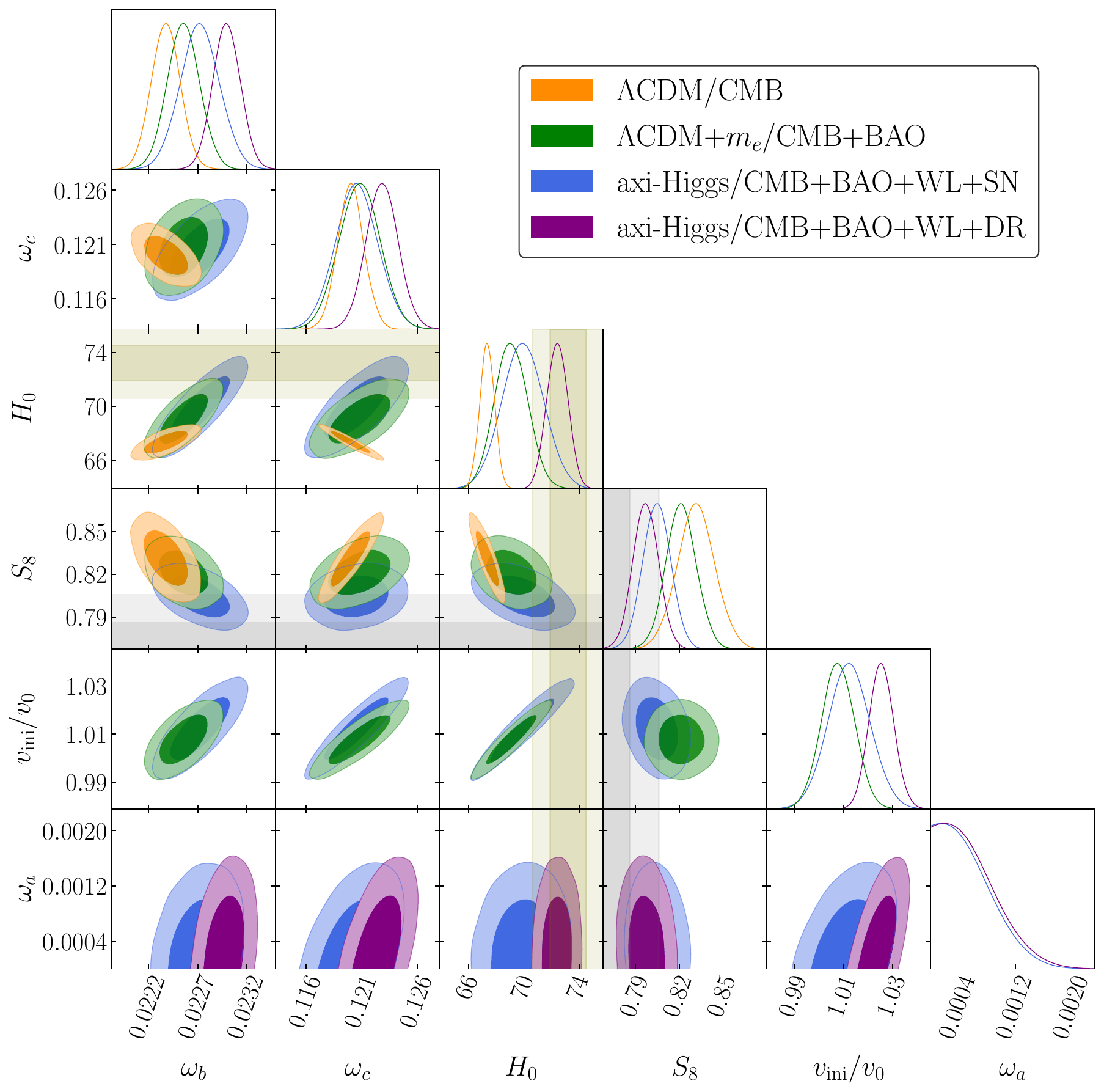}
		\caption{Posterior distributions of the model parameters in the axi-Higgs benchmark scenario. The shaded olive and grey bands represent the local DR measurement of $H_0$ from the latest SH0ES-20~\cite{Riess:2020fzl} and the weak-lensing measurement of $S_8$ from KiDS-1000~\cite{Heymans:2020gsg}, respectively.}
		\label{fig:model_comparison}
	\end{figure}

We demonstrate the axi-Higgs posterior distributions for the benchmark of $m=2 \times 10^{-30}$ eV in Fig.~\ref{fig:model_comparison} (see supplemental material at Sec. E for the impacts of the axion mass on data fitting, and at Sec. F for an overall picture on the axi-Higgs cosmology). Compared to the $\Lambda$CDM/CMB~\cite{Aghanim:2018eyx} and $\Lambda$CDM$+m_e$/CMB$+$BAO~\cite{Hart:2019dxi} analyses, the axi-Higgs/CMB$+$BAO$+$WL+SN scenario yields a higher $H_0$ value, accompanied by a downward shift of $S_8$ (due to $S_{8|a} \ll 0$ and $\omega_a > 0$). The blue filled contours overlap with the intersection of the shaded olive and grey bands in the $H_0$-$S_8$ panel. The $H_0$ and $S_8$ tensions are thus simultaneously reduced!

\section{Summary and Remarks}

As a low-energy effective theory motivated by string theory, the axi-Higgs model broadly impacts our understanding on the universe~\cite{Fung:2021wbz}. In this paper, we have demonstrated how the $H_0$ and $S_8$ tensions get simultaneously relaxed in this model, by correlating the axion impacts on the early and late universe.

In the early universe ($z>z_a$), this axion field behaves like DE. Its main impact is to drive a positive shift in the Higgs VEV. In the late universe ($z<z_a$),  this axion field behaves like DM. Its main impacts are to: (1) increase the $H(z)$ value during this epoch and hence reduce the comoving diameter distance at $z_*$ and $z_{\rm eff}$; (2) suppress the formation of the structure at a galactic clustering scale and even above; and (3) shift the $v$ (or $m_e$) value in the early universe to its today's value $v_0$. Combining the axion impact at $z>z_a$ and item (1) mitigates the Hubble tension, further including item (2) relaxes the $S_8$ tension, and finally including item (3) restores our observation on today's electron.     

To conclude, we stress that a full test of this model is at hand, due to the oncoming AC and the quasar spectral measurements with the data expected to be collected by, e.g., Thirty Meter Telescope~\cite{Skidmore:2015lga} and James Webb Space Telescope~\cite{Behroozi:2020jhj}. More details on this can be found in~\cite{Fung:2021wbz}.

\section{Acknowledgement}

We thank Luke Hart and Jens Chluba for valuable communications. This work is supported partly by the Area of Excellence under the Grant No.~AoE/P-404/18-3(6), partly by the General Research Fund under Grant No.~16305219, and partly by the Collaborative Research Fund under the Grant No.~C7015-19G.  All grants were issued by the Research Grants Council of Hong Kong S.A.R.

\section{Supplemental Materials}

The Supplementary Materials contain additional calculations and analyses in support of the results presented in this paper. In Sec.~A, we make a pedagogical introduction to the set of compressed cosmological observables (including the relevant data) which are applied in this study. In Sec.~B, we present the detailed derivation of variation equations for this set of compressed observables. We demonstrate the dependence of the axion perturbation effects on its mass in Sec.~C, and provide the details of testing the LPA validity in the $\Lambda$CDM, $\Lambda$CDM$+m_e$ and axi-Higgs models in Sec.~D. The impacts of the axion mass on the fitting results in the axi-Higgs model are discussed in Sec.~E. Finally, we present an overall picture on the axi-Higgs cosmology in Sec.~F, to further highlight the significance of the study in this Letter.

\section*{A. Cosmological observables} \label{App:first}

In the LPA analysis in~\cite{Fung:2021wbz}, we consider only the angular sound horizon at recombination $\theta_*$ and the quantity $r_d h$ as the CMB and BAO observables  respectively, for a simple demonstration. In this study, we take a much more comprehensive treatment, by including the CMB scale parameters ($l_a$, $l_{\rm eq}$ and $l_D$) and $S_L=\sigma_8\Omega_m^{0.25}$, the BAO scale parameters ($\alpha_V$, $\alpha_\perp$ and $\alpha_\parallel$), the $S_8$ parameter and the supernova $m_B$ parameter (or the local DR measurements). The data, reference value and covariance matrix for these observables are summarized in Tab.~\ref{tab:data}.

\subsection{CMB angular horizons}
The three scale parameters of CMB include~\cite{Hu:2000ti}:
\begin{itemize}
	\item Sound horizon at recombination
	\begin{eqnarray}
	l_a \equiv \dfrac{\pi}{\theta_*} = \pi \dfrac{D_*}{r_*} \; .
	\end{eqnarray}
	Here $r_* \equiv r_s(z_*)$ and $D_* \equiv D_M(z_*)$ are sound horizon at recombination and diameter distance from recombination, with
	\begin{eqnarray}
	r_s(z) &=&  \int_{z}^{\infty} dz'\dfrac{c_s(z')}{H(z')} \ ,  \\
	c_s(z) &=&  \frac{1}{ \sqrt{3}} \left ( 1 + \dfrac{3\rho_b(z)}{4\rho_\gamma(z)} \right)^{-\frac{1}{2}}  \ , 
	\end{eqnarray} 
	and
	\begin{eqnarray} D_M (z) =\int_{0}^{z} \dfrac{dz'}{H(z')} \ .
	\end{eqnarray} 
	Both of them are comoving. As in \texttt{CAMB}~\cite{Lewis:1999bs} and also in~\cite{ Hu:1995en}, the recombination redshift $z_*$ is defined as the moment at which the optical depth $\tau (z)$ reaches unity, namely $\tau(z_*) = 1$, with
	\begin{align}
	\tau (z) = \int^z_0 dz \dfrac{\sigma_T n_e (z)}{(1 + z) H(z)} \ .
	\end{align}
	
	\item Hubble horizon at matter-radiation equality
	\begin{eqnarray}
	l_\text{eq} \equiv k_\text{eq} D_*   = \dfrac{H(z_\text{eq})}{1 + z_\text{eq}} D_* \ .
	\end{eqnarray}  
	Here $z_\text{eq}$ is determined by 
	\begin{eqnarray}
	\Omega_m (z_\text{eq}) = \Omega_r (z_\text{eq}) \ , 
	\end{eqnarray}
	with 
	\begin{eqnarray} 
	\Omega_m(z_\text{eq}) &=& (\Omega_b^0 + \Omega_c^0)(1 + z_\text{eq})^3 + \Omega_a(z_\text{eq}) \;,   \\
	\Omega_r(z_\text{eq}) &=& \Omega_r^0(1 + z_\text{eq})^4 + \Omega_\nu(z_\text{eq}) \ . 
	\end{eqnarray}
	
	\item Damping scale at recombination
	\begin{eqnarray}
	l_D \equiv k_D D_* \ , 
	\end{eqnarray}  
	with
	\begin{eqnarray} k_D =  \left( \dfrac{1}{6} \int_{z_*}^{\infty} \dfrac{dz}{H(z) \tau'} \dfrac{R^2 + 16 (1 + R) / 15}{(1 + R)^2} \right)^{-\frac{1}{2}}  \ .
	\end{eqnarray}
	Here the differential of the optical depth and baryon-photon ratio are given by 
	\begin{eqnarray}
	\tau' \equiv \dfrac{\sigma_T n_e(z)}{1 + z}, \qquad R (z) \equiv \dfrac{3 \omega_b}{4 \omega_\gamma} \frac{1}{1 + z}\; .
	\end{eqnarray}
	
\end{itemize}

\subsection{CMB lensing and $S_L$}

The CMB lensing power spectrum $C^{\rm{\phi\phi}}_{\ell}$ can be largely encapsulated by~\cite{Planck:2018lbu}  
\begin{eqnarray}
S_L \equiv \sigma_8 \Omega_m^{0.25}  \ .
\end{eqnarray}
This parameter combination can be understood as follows. The CMB lensing spectrum is a convolution of the CMB power spectrum and the integrated foreground matter power spectrum responsible for lensing. The two power spectra are each characterised by $\sigma_8$, and the growth of structure relevant for integrating the line of sight lens modify the power at a rate of $\Omega_m^{0.5}$. The result of such convolution therefore give rises to $\sigma_8^2\Omega_m^{0.5}$. By convention we take the square root, which yields the observable $S_L$ similar to but different from $S_8$. 

In this study, we apply the data of Planck 2018~\cite{Aghanim:2018eyx} to define the observation values of ${\bf Y_{\rm CMB}} = \{l_a, l_\text{eq}, l_D, S_L \}$. 
Its covariance matrix is then given by
\begin{eqnarray} \label{Eq:cmb_cov}
\mathbf{\Sigma}_{\rm CMB} 
= \begin{pmatrix}
0.0079 &  0.0287  & -0.0099 & 0.0001\\
0.0287 &  0.7917  & -0.5164 & 0.0043\\
-0.0099 & -0.5164  &  4.5656 & -0.0036\\
0.0001 & 0.0043 & -0.0036 & 0.00004
\end{pmatrix}  \ .
\end{eqnarray}	
Here 
\begin{eqnarray}
\mathbf{\Sigma}_{ij} &=& \rho_{ij}\sigma_i\sigma_j = \frac {\sum_{k=1}^n (Y_{i,k}- {\bar Y}_i)(Y_{j,k}- {\bar Y}_j)}{n - 1}
\end{eqnarray}
has been used to calculate the matrix elements, with $n$ being the number of data points from each sample and ${\bar Y}_i$ and ${\bar Y}_j$ being their respective means.

\subsection{BAO sound horizons}

The decoupling of baryons from photons freezes their fluctuations. It then leaves an imprint at the $r_d$ scale in the matter power spectrum which can be probed by the large-scale-structure surveys. Here $r_d = r_s (z_d)$ is the sound horizon at the end of baryon drag. The baryon-drag redshift $z_d$ is defined as the moment at which the baryon-drag depth $\tau_d (z)$ reaches unity, namely $\tau_d (z_d)=1$, with
\begin{align}
\tau_d(z) = \int^z_0 \dfrac{d\tau/dz}{R (z)} \ .
\end{align}

Similar to the CMB case, the matter power spectrum can be characterized by three scale parameters:
\begin{itemize}
	\item BAO scale perpendicular to the line-of-sight 
	\begin{eqnarray}
	\alpha_\perp (z_\text{eff}) \propto \dfrac{D_M (z_\text{eff})}{r_d} \equiv \dfrac{(1 + z_\text{eff}) D_A(z_\text{eff})}{r_d} \; .
	\end{eqnarray}
	\item BAO scale parallel to the line-of-sight
	\begin{eqnarray}
	\alpha_\parallel (z_\text{eff}) \propto \dfrac{D_H(z_\text{eff})}{r_d} \equiv \dfrac{1}{H(z_\text{eff}) r_d} \; .
	\end{eqnarray}
	\item Isotropic BAO scale
	\begin{eqnarray}
	\alpha_V (z_\text{eff}) \propto \dfrac{D_V(z_\text{eff})}{r_d} \equiv \left ( \dfrac{ z_\text{eff} D_M(z_\text{eff})^2 }{ H(z_\text{eff})} \right )^{\frac{1}{3}} \frac{1}{r_d} \; .
	\end{eqnarray}
\end{itemize}
These BAO scale parameters are usually measured at an effective redshift $z_\text{eff}$. 

In this study, we combine the BAO data in two different ways and call them with two separate names to avoid confusion:
\begin{itemize}
	\item DR12: To test the LPA validity in the $\Lambda$CDM$+m_e$ model (see Sec.~D), we apply the BOSS DR12 data including 6dF at $z_\text{eff} = 0.106$, MGS at $z_\text{eff} = 0.15$, LOWZ at $z_\text{eff} = 0.32$, and CMASS at $z_\text{eff} = 0.57$, the same as the ones used in~\cite{Hart:2019dxi}.
	\item BAO: To analyze the cosmological parameters in the axi-Higgs model (see the main text), we opt for the most updated eBOSS data including 6dF at $z_\text{eff} = 0.106$, MGS at $z_\text{eff} = 0.15$, LRG at $z_\text{eff} = 0.698$, ELG at $z_\text{eff} = 0.845$, Quasar at $z_\text{eff} = 1.48$, and Lyman-$\alpha$ at $z_\text{eff} = 2.334$.
\end{itemize}
We will use the ``DR12'' data to test the LPA validity in $\Lambda$CDM+$m_e$ (the same data have been applied for the $\Lambda$CDM+$m_e$ study in~\cite{Hart:2019dxi}), and the ``BAO'' data for the other analyses. The mean of these BAO observables can be found in Tab.~\ref{tab:data}. Their covariance matrices are given by~\cite{Beutler_2011, Ross:2014qpa, Gil-Marin:2015nqa, Bautista:2020ahg, Raichoor:2020vio, Neveux:2020voa, duMasdesBourboux:2020pck}
\begin{widetext}
	\begin{eqnarray}
	\mathbf{\Sigma}_{\rm DR12}
	&=& \begin{pmatrix}
	0.0193 & 0 & 0 & 0 & 0 & 0 \\
	0 & 0.0283 & 0 & 0 & 0 & 0 \\
	0 & 0 & 0.0225 & 0.0362 & 0 & 0 \\
	0 & 0 & 0.0362 & 0.4761 & 0 & 0 \\
	0 & 0 & 0 & 0 & 0.0144 & 0.0262 \\
	0 & 0 & 0 & 0 & 0.0262 & 0.1764
	\end{pmatrix}  \ ,  \label{Eq:dr12_cov}  \\
	\mathbf{\Sigma}_{\rm BAO}
	&=& \begin{pmatrix}
	0.0193 & 0 & 0 & 0 & 0 & 0 & 0 & 0 & 0 \\
	0 & 0.0283 & 0 & 0 & 0 & 0 & 0 & 0 & 0 \\
	0 & 0 & 0.111 & -0.0586 & 0 & 0 & 0 & 0 & 0 \\
	0 & 0 & -0.0586 & 0.292 & 0 & 0 & 0 & 0 & 0 \\
	0 & 0 & 0 & 0 & 0.3364 & 0 & 0 & 0 & 0 \\
	0 & 0 & 0 & 0 & 0 & 0.6373 & 0.1707 & 0 & 0 \\
	0 & 0 & 0 & 0 & 0 & 0.1707 & 0.3047 & 0 & 0 \\
	0 & 0 & 0 & 0 & 0 & 0 & 0 & 1.44 & -0.108 \\
	0 & 0 & 0 & 0 & 0 & 0 & 0 & -0.108 & 0.04 \\
	\end{pmatrix} \ .  \label{Eq:bao_cov}
	\end{eqnarray} 
\end{widetext}

\subsection{Matter clustering amplitude}

The fluctuation amplitude of matter density at the scale of $8h^{-1}$ Mpc (a typical scale for galactic clusters) can be probed by the galaxy-clustering observations and the weak-lensing experiments. Its observable is conventionally defined as
\begin{eqnarray}
S_8 \equiv \sigma_8 \left( \dfrac{\Omega_m}{0.3} \right)^{0.5} \; .
\end{eqnarray}
In the likelihood analysis, the WL data is encoded as the split-normal priors of $S_8$, $i.e.$~\cite{Abbott:2017wau, Heymans:2020gsg}, 
\begin{align}
\mathbf{\Sigma}_{\rm DES} = 
\begin{cases}
0.676 \times 10^{-3} \quad (S_8 \geq \mathbf{Y_{\rm o,DES}}) \\
0.4 \times 10^{-3} \quad \ \ \  (S_8 < \mathbf{Y_{\rm o,DES}})
\end{cases} \ , \\
\mathbf{\Sigma}_{\rm KiDs} = 
\begin{cases}
0.4 \times 10^{-3} \quad   \ \ \  (S_8 \geq \mathbf{Y_{\rm o,KiDs}}) \\
0.196 \times 10^{-3} \quad (S_8 < \mathbf{Y_{\rm o,KiDs}})
\end{cases} \ .
\end{align}
The total WL covariance matrix is then given by
\begin{eqnarray}
\mathbf{\Sigma}_{\rm WL}
&=& \begin{pmatrix}
\mathbf{\Sigma}_{\rm DES} & 0 \\
0 & \mathbf{\Sigma}_{\rm KiDs}
\end{pmatrix} \ . \label{Eq:wl_cov}
\end{eqnarray}

\subsection{Supernova luminosity and DR measurements}

The flux measurements of low-redshift supernovae provide the information on the Hubble flow of the late-time universe. Here the apparent luminosity is defined as~\cite{Scolnic:2017caz}
\begin{eqnarray}
m_B \equiv \mu + M_0 = 5\log_{10}\left(\dfrac{D_L}{{\rm Mpc}}\right) + 25 + M_0 \ .
\end{eqnarray}
$\mu$ is distance modulus, with $D_L (z_{\rm eff}) = (1+z_{\rm eff}) D_M (z_{\rm eff})$ being luminosity distance, and $M_0$ is absolute luminosity. 
$M_0$ is a quantity yet to be determined. Usually it is calculated as a weighted average of the apparent-luminosity data at each sampling step in the MCMC chain, namely
\begin{align}
M_0 = \sum_i \left(m_{B, i} - \mu_i(\mathbf{X}) \right) \sigma^{-2}_i / \sum_i \sigma^{-2}_i  \ .
\end{align}

In this study, we compress the original data with 1048 SNs~\cite{Scolnic:2017caz} into four redshift bins, using BEAMS with bias corrections or BBC method~\cite{Kessler:2016uwi}. Then the index $i$ runs over four data points each of which is defined by one redshift bin. For each data point, $m_{B,i}$ and $\sigma_i$ denote the mean of apparent luminosity, see Tab.~\ref{tab:data}, and its statistical uncertainty, respectively. The SN measurements also subject to systematic uncertainties, so we have the total covariance matrix 
\begin{align}
\mathbf{\Sigma}_{\rm SN} = \mathbf{\Sigma}_{\rm stat} + \mathbf{\Sigma}_{\rm sys} \ .
\end{align}
Here $\mathbf{\Sigma}_{\rm stat}$ is a diagonal matrix with its diagonal entries defined by $\sigma^2_i$, whereas $\mathbf{\Sigma}_{\rm sys}$ is obtained as an interpolation of the original $40\times 40$ covariance matrix presented in~\cite{Scolnic:2017caz}. Then they are given by
\begin{eqnarray}
\mathbf{\Sigma}_{\rm stat}
= 10^{-4} \begin{pmatrix}
0.893 & 0 & 0 & 0 \\
0 & 0.648 & 0 & 0 \\
0 & 0  &  0.497 & 0 \\
0 & 0 & 0 & 0.837
\end{pmatrix}  \ ,  \label{Eq:sn_stat_cov} \\
\mathbf{\Sigma}_{\rm SN} 
= 10^{-4} \begin{pmatrix}
7.51 & -0.605 & -0.261 & -1.277 \\
-0.605 & 1.443 & 0.278 & -1.977 \\
-0.261 & 0.278  & 0.86 & -1.716 \\
-1.277 & -1.977 & -1.716 & 27.65
\end{pmatrix} \ . \label{Eq:sn_cov}
\end{eqnarray}

Although the SN constraints on the late-time Hubble flow (see, e.g.~\cite{Dainotti:2021pqg}) and the DR determination of the $H_0$ value are both based on the SN data, significant difference exists between them. To pin down the $H_0$ value, we need to properly calibrate the $M_0$ value of the supernovae using  independent sources, while this is not necessary for constraining the Hubble flow. In this paper, we refer to ``SN'' as the uncalibrated data, while ``DR'' as the data calibrated with local distance ladders. The DR data is then implemented with Gaussian priors of $H_0$ which is inferred from a series of late-time observations. Following Ref.~\cite{Verde:2019ivm}, we assume the correlations between these measurement to be weak. The covariance matrix is then given by~\cite{Riess:2019cxk, Wong:2019kwg, Reid:2008nm, Freedman:2019jwv, Potter:2018, Huang:2018dbn}
\begin{eqnarray}
\mathbf{\Sigma}_{\rm DR}
= \begin{pmatrix}
1.96 & 0 & 0 & 0 & 0 & 0 \\
0 & 3.24 & 0 & 0 & 0 & 0 \\
0 & 0 & 9.61 & 0 & 0 & 0 \\
0 & 0 & 0 & 3.61 & 0 & 0 \\
0 & 0 & 0 & 0 & 16 & 0 \\
0 & 0 & 0 & 0 & 0 & 15.21
\end{pmatrix} .  \label{Eq:dr_cov}
\end{eqnarray}

\begin{table}[htp]
	\resizebox{0.48\textwidth}{!}{
		\centering
		\begin{tabular}{ |c|c|c|c|c|c| }
			\hline
			Data & Experiments & Observables & Reference (${\rm Y}_{\rm ref}$) & Mean (${\rm Y}_{\rm o}$) & Correlation ($\Sigma$) \\
			\hline 
			\multirow{4}{*}{CMB} & \multirow{4}{2.3cm}{Planck 2018~\cite{Aghanim:2018eyx} (TT,TE,EE +lowE+lensing)}& $l_a$ & $301.765$ & $301.757$ & \multirow{4}{*}{Eq.~\eqref{Eq:cmb_cov}}  \\
			\cline{3-5}
			& & $l_\text{eq}$ & $144.14$ & $144.05$ &  \\
			\cline{3-5}
			& & $l_D$ & $1954.6$ & $1954.3$ &  \\
			\cline{3-5}
			& & $S_L$ & $0.6085$ & $0.6078$ &  \\
			\hline
			\multirow{6}{*}{DR12} & 6dF~\cite{Beutler_2011} & $D_V(0.106)/r_d$ & $3.10$ & $2.98$ & \multirow{6}{*}{Eq.~\eqref{Eq:dr12_cov}} \\
			\cline{2-5}
			& SDSS MGS~\cite{Ross:2014qpa} & $D_V(0.15)/r_d$ & $4.32$ & $4.47$ & \\
			\cline{2-5}
			& \multirow{2}{*}{BOSS LOWZ~\cite{Gil-Marin:2015nqa}} & $D_A(0.32)/r_d$ & $6.76$ & $6.67$ &  \\
			\cline{3-5}
			& & $H(0.32)r_d$ & $11.76$ & $11.63$ &  \\
			\cline{2-5}
			& \multirow{2}{*}{BOSS CMASS~\cite{Gil-Marin:2015nqa}} & $D_A(0.57)/r_d$ & $9.46$ & $9.47$ & \\
			\cline{3-5}
			& & $H(0.57)r_d$ & $13.67$ & $14.67$ &  \\
			\hline
			\multirow{9}{*}{BAO} & 6dF~\cite{Beutler_2011} & $D_V(0.106)/r_d$ & $3.10$ & $2.98$ & \multirow{9}{*}{Eq.~\eqref{Eq:bao_cov}} \\
			\cline{2-5}
			& SDSS MGS~\cite{Ross:2014qpa} & $D_V(0.15)/r_d$ & $4.32$ & $4.47$ & \\
			\cline{2-5}
			& \multirow{2}{*}{eBOSS LRG~\cite{Bautista:2020ahg}} & $D_M(0.698)/r_d$ & $17.55$ & $17.86$ & \\
			\cline{3-5}
			& & $D_H(0.698)/r_d$ & $20.28$ & $19.34$ & \\
			\cline{2-5}
			& eBOSS ELG~\cite{Raichoor:2020vio} & $D_V(0.845)/r_d$ & $18.68$ & $18.23$ & \\
			\cline{2-5}
			& \multirow{2}{*}{eBOSS Quasar~\cite{Neveux:2020voa}} & $D_M(1.48)/r_d$ & $30.24$ & $30.69$ & \\
			\cline{3-5}
			& & $D_H(1.48)/r_d$ & $12.91$ & $13.26$ & \\
			\cline{2-5}
			& \multirow{2}{*}{eBOSS Ly-$\alpha$~\cite{duMasdesBourboux:2020pck}} & $D_M(2.334)/r_d$ & $39.23$ & $37.5$ & \\
			\cline{3-5}
			& & $D_H(2.334)/r_d$ & $8.60$ & $8.99$ & \\
			\hline
			\multirow{2}{*}{WL} & DES Y1 3x2~\cite{Abbott:2017wau} & \multirow{2}{*}{$S_8$} & \multirow{2}{*}{$0.833$} & $0.773$ & \multirow{2}{*}{Eq.~\eqref{Eq:wl_cov}} \\
			\cline{2-2}\cline{5-5}
			& KiDS-1000~\cite{Heymans:2020gsg} &  &  & $0.766$ &  \\
			\hline
			\multirow{4}{*}{SN} & \multirow{4}{2.3cm}{Pantheon~\cite{Scolnic:2017caz} (binned)} & $m_B(0.047)$ & $17.2843$ & $17.2474$ & \multirow{4}{*}{Eq.~\eqref{Eq:sn_cov}} \\
			\cline{3-5}
			&  & $m_B(0.142)$ & $19.8017$ & $19.8065$ & \\
			\cline{3-5}
			& & $m_B(0.309)$ & $21.6978$ & $21.6928$ & \\
			\cline{3-5}
			&  & $m_B(1.36)$ & $25.5768$ & $25.6154$ & \\
			\hline
			\multirow{6}{*}{DR} & SH0ES-19~\cite{Riess:2019cxk} & \multirow{6}{*}{} & \multirow{6}{*}{$67.32$} & $74.0$ & \multirow{6}{*}{Eq.~\eqref{Eq:dr_cov}} \\
			\cline{2-2}\cline{5-5}
			& H0LiCOW~\cite{Wong:2019kwg} &  &  & $73.3$ & \\
			\cline{2-2}\cline{5-5}
			& MCP~\cite{Reid:2008nm} & $H_0$ &  & $74.8$ & \\
			\cline{2-2}\cline{5-5}
			& CCHP~\cite{Freedman:2019jwv} & (km/s/Mpc) &  & $69.8$ & \\
			\cline{2-2}\cline{5-5}
			& SBF~\cite{Potter:2018} &  &  & $76.5$ & \\
			\cline{2-2}\cline{5-5}
			& MIRAS~\cite{Huang:2018dbn} &  &  & $73.6$ & \\
			\hline
	\end{tabular}		}
	\caption{Data and reference values for the observables applied in the LPA analysis.}
	\label{tab:data}
\end{table}

\subsection{Evolution of the axi-Higgs axion}

Neglecting its feedback on Hubble flow, the axion evolution can be solved from the equation 
\begin{align}
\ddot{\theta} + 3 H\dot{\theta} + m_1^2 \theta = 0 \ .
\end{align}
Here $\theta = a_1 / a_{1,\text{ini}}$ satisfies the initial conditions $\theta_\text{ini} = 1$ and $\dot{\theta}_\text{ini} = (d\theta / dt)_\text{ini} = 0$ is its initial derivative w.r.t cosmic time. Then we have 
\begin{eqnarray}
\Omega_a(z) = \Omega_a^0 g(z), 
\end{eqnarray}
with 
\begin{align}
g(z) = \dfrac{1}{2\theta_0^2} \left( \dfrac{\dot{\theta}^2}{m_1^2}  + \theta^2 \right) , \quad \theta_0 = \theta(z=0) \; .
\end{align}
This provides a more complete form of Eq.~(\ref{eq:a1osc}).

\section*{B. Variation Equations in the axi-Higgs Model}\label{App:second}

Varying the cosmological observables introduced in Sec.~A yields
\begin{eqnarray}
\delta l_a &=& \delta D_* - \delta r_* \; , \label{eq:cmb_1} \\
\delta l_\text{eq} &=& \delta k_\text{eq} + \delta D_*  \nonumber \\
&=& \delta H_\text{eq} - \dfrac{z_\text{eq}}{1 + z_\text{eq}} \delta z_\text{eq} + \delta D_* \; , \label{eq:cmb_2} \\
\delta l_D &=& \delta k_D + \delta D_* \;  , \\
\delta S_L &=& \delta \sigma_8 + 0.25 \delta \Omega_m \; ,   \\
\delta \alpha_\perp (z_\text{eff}) &=& \delta D_M (z_\text{eff}) - \delta r_d \; , \label{eq:bao_1} \\
\delta \alpha_\parallel (z_\text{eff}) &=& - \delta H (z_\text{eff}) - \delta r_d \; , \label{eq:bao_2} \\
\delta \alpha_V (z_\text{eff}) &=& \dfrac{2}{3} \delta D_M(z_\text{eff}) - \dfrac{1}{3} \delta H (z_\text{eff})  - \delta r_d \; ,   \\
\delta S_8 &=& \delta \sigma_8 + 0.5 \delta \Omega_m \; ,   \\
\delta \mu (z_{\rm eff}) &=& \dfrac{d\mu(z_{\rm eff})}{\mu(z_{\rm eff})} = \dfrac{5\log_{10}e}{\mu(z_{\rm eff})} \delta D_M (z_{\rm eff}) \; , \\
\delta m_B (z_{\rm eff}) &=& \dfrac{ d\mu(z_{\rm eff}) - \sum_i \left( d\mu_i \sigma^{-2}_i \right) / \sum_i \sigma^{-2}_i }{\mu(z_{\rm eff}) + M_0} \; ,
\end{eqnarray}
with			
\begin{eqnarray}
\delta\Omega_m &=& \dfrac{\omega_b}{\omega_m} \delta\omega_b + \dfrac{\omega_c}{\omega_m} \delta\omega_c - 2 \delta h + \dfrac{\omega_a}{\omega_m}  \ .
\end{eqnarray}
We then calculate the partial derivatives of their intermediate quantities w.r.t. the axi-Higgs parameters, 
\begin{eqnarray}   
r_{s|b} &=& - \dfrac{\mathcal{C} \omega_b}{2r_s}  \int_{z}^{\infty} dz' \dfrac{c_s(z')}{h(z')} \nonumber \\
&& \times \left[ \dfrac{9}{4} \dfrac{c_s^2(z')}{\omega_\gamma (1 + z')} + \dfrac{(1 + z')^3 - 1}{h^2(z')} \right]\ , \\
r_{s|c} &=& -\dfrac{\mathcal{C} \omega_c}{2r_s} \int_{z}^{\infty} dz' \dfrac{c_s(z')}{h^3(z')} \left[ (1+z')^3 - 1 \right]\ , \\
r_{s|h} &=& -\dfrac{\mathcal{C} h^2}{r_s} \int_{z}^{\infty} dz' \dfrac{c_s(z')}{h^3(z')}\ , \\ 
r_{s|z} &=& - \dfrac{\mathcal{C}z}{r_s} \dfrac{c_s(z)}{h(z)} \ , \\
r_{s|a} &=& -\dfrac{\mathcal{C}}{2r_s} \int_{z}^{\infty} dz' \dfrac{c_s(z')}{h^3(z')} \left[ g(z') - 1 \right] \ , \\
D_{M|b,c} &=& - \dfrac{\mathcal{C} \omega_{b,c}}{2D_M}  \int^{z}_{0} \dfrac{dz'}{h^3(z')} \left[ (1+z')^3 - 1 \right]\ , \\
D_{M|h} &=& - \dfrac{\mathcal{C} h^2}{D_M}  \int^{z}_{0} \dfrac{dz'}{h^3(z')}\ , \\ 
D_{M|z} &=& \dfrac{\mathcal{C} z}{D_M(z) h(z)} \ \ , \\
D_{M|a} &=& - \dfrac{\mathcal{C}}{2D_M}  \int^{z}_{0} \dfrac{dz'}{h^3(z')} \left[ g(z') - 1 \right]\ , \\
H_{|b,c} &=& \dfrac{\omega_{b,c}}{2h(z)^2} \left[ (1+z)^3 - 1 \right]\ , \\ 
H_{|h} &=& \dfrac{h^2}{h(z)^2} \ , \\
H_{|a} &=& \dfrac{1}{2h(z)^2} [ g(z) - 1 ] \ ,
\end{eqnarray}
with $\mathcal{C} = 2998$ Mpc. Notice that $\omega_\Lambda = h^2 - \omega_b -\omega_c -\omega_a$ due to flatness requirement. Notably, if the physical quantities are defined at some specific redshift, their partial derivatives also receive a contribution from the redshift change caused by the parameter variations, namely  
\begin{eqnarray}
r_{*|x} &=& r_{s|x}(z_*) + r_{s|z}(z_*) z_{*|x} \ , \\
r_{d|x} &=& r_{s|x}(z_d) + r_{s|z}(z_d) z_{d|x} \ , \\
D_{*|x} &=& D_{M|x}(z_*) + D_{M|z}(z_*) z_{*|x} \ , \\
H (z_\text{eq})_{|x} &=& H_{|x}(z_\text{eq}) + H_{|z}(z_\text{eq}) z_{\text{eq}|x} \ ,
\end{eqnarray}
with $x = a,b,c, h, v$. At last, we calculate the derivatives of the redshifts $z_{*}$, $z_d$ and $z_\text{eq}$, together with $H_{|z}(z_\text{eq})$ and $k_{D|x}$, numerically, using the center second-order formula
\begin{eqnarray}
Y_{|x} = \dfrac{Y(1.001x) - Y(0.999x)}{0.002 Y(x)}
\end{eqnarray}
for $x=b,c,v,h$ and the one-sided formula 
\begin{eqnarray}
Y_{|a} = \dfrac{-3Y(0) + 4Y(0.001) - Y(0.002)}{0.002 Y(0)}
\end{eqnarray}
for $\omega_a$. Here \texttt{Recfast++/HyRec}~\cite{Chluba:2010ca, Ali_Ha_moud_2011, Lee:2020obi} are applied to calculate $\tau'$ (more specifically $n_e(z)$), with a recombination history modified by $\delta v$. \texttt{CAMB/CLASS}~\cite{Lewis:1999bs, Blas_2011} and \texttt{axionCAMB}~\cite{Hlozek:2014lca, Hlozek:2017zzf} are also applied for calculating $\sigma_{8|b,c,h,v}$ and $\sigma_{8|a}$, respectively. We notice that in \texttt{axionCAMB} the axion is identified as matter in terms of its contribution to the energy density. This is not an accurate treatment for the ultra-light axions such as $a$, especially for the calculation of $z_\text{eq}$. The numerical values of $Y_{|X}$ for the observables and the intermediate quantities in this study are summarized in Tab.~\ref{tab:par_derivs}.

\begin{table}[htp]
	\resizebox{0.46\textwidth}{!}{
		\centering
		\begin{tabular}{ |c|c|c|c|c|c| }
			\hline
			$Y \backslash  X$  & $ \omega_b$ & $ \omega_c$ &  $ h$ & $ v_{\rm ini}$ & $\omega_a$  \\	
			\hline 			
			\hline
			$ l_a$ & $0.0550$ & $- 0.1203$ & $- 0.1934$ &  $0.6837$ & $- 2.6364$ \\		
			\hline
			$ l_\text{eq}$ & $0.0942$ & $0.5082$ & $-0.1934$ & $0.0154$ & $- 2.6376$   \\
			\hline
			$ l_D$ & $0.2459$ & $- 0.0962$ & $- 0.1934$ & $0.4553$ & $-2.6315$ \\		
			\hline
			$S_L$ & $-0.1398$ & $0.8484$ & $-0.2619$ & $0.0788$  &  $-16.082$  \\		
			\hline
			$ \alpha_V (0.106)$ & $0.1635$ &  $0.1874$ & $- 0.9334$ &  \multirow{13}{*}{$0.6162$} & $-0.2312$ \\ \cline{1-4}\cline{6-6}
			$ \alpha_V (0.15)$ & $0.1613$ &  $0.1760$ & $-0.9062$ &  & $-0.3261$ \\ \cline{1-4}\cline{6-6} 
			$ \alpha_\perp$(0.32) & $0.1574$ & $0.1547$ & $-0.8554$ &  & $-0.5037$  \\
			\cline{1-4}\cline{6-6}
			$ \alpha_\parallel$(0.32) & $0.1459$ &$0.0933$ & $-0.7089$ &  & $-1.0152$  \\
			\cline{1-4}\cline{6-6}
			$ \alpha_\perp$(0.57) & $0.1499$ & $0.1145$ & $-0.7596$ &  & $-0.8382$  \\
			\cline{1-4}\cline{6-6}
			$ \alpha_\parallel$(0.57) & $0.1315$ &$0.0159$ & $-0.5245$ &  & $-1.6591$  \\
			\cline{1-4}\cline{6-6}
			$ \alpha_\perp$(0.698) & $0.1466$ & $0.0969$ & $-0.7175$ &  & $-0.9852$  \\
			\cline{1-4}\cline{6-6}
			$ \alpha_\parallel$(0.698) & $0.1256$ &$-0.0160$ & $-0.4483$ &  & $-1.9252$  \\
			\cline{1-4}\cline{6-6}
			$ \alpha_V$(0.845) & $0.1354$ &  $0.0370$ & $-0.5747$ &  & $-1.4838$  \\
			\cline{1-4}\cline{6-6}
			$ \alpha_\perp$(1.48) & $0.1330$ & $0.0238$ & $-0.5431$ & & $-1.5940$  \\
			\cline{1-4}\cline{6-6}
			$ \alpha_\parallel$(1.48) & $0.1047$ & $-0.1278$ & $-0.1817$ &  & $-2.8556$  \\
			\cline{1-4}\cline{6-6}
			$ \alpha_\perp$(2.334) & $0.1255$ & $-0.0163$ & $-0.4475$ &  & $-1.9277$  \\
			\cline{1-4}\cline{6-6}
			$ \alpha_\parallel$(2.334) & $0.0969$ & $-0.1700$ & $-0.0807$ &  & $-3.2076$  \\
			\hline 
			$ S_8$ & $-0.1007$ & $1.0581$ & $-0.7619$ & $0.0788$  &  $-14.335$  \\
			\hline
			$ m_B (0.047)$ & $0.0013$ & $0.0072$ & $-0.0171$ & \multirow{5}{*}{$0$}  &  $0.06$  \\
			\cline{1-4}\cline{6-6}
			$ m_B (0.142)$ & $0.0008$ & $0.0043$ & $-0.0101$ &  &  $0.0355$  \\
			\cline{1-4}\cline{6-6}
			$ m_B (0.309)$ & $0.0001$ & $0.0008$ & $-0.0019$ &  &  $0.0066$  \\
			\cline{1-4}\cline{6-6}
			$ m_B (1.36)$ & $-0.0018$ & $-0.0099$ & $0.0237$ &  &  $-0.0829$  \\
			\hline   
			\hline
			$z_*$ & $-0.0264$ & $0.0097$ & \multirow{2}{*}{$0$} & $1.0184$ & $0.0011$  \\
			\cline{1-3}\cline{5-6}
			$r_*$ & $-0.1178$ & $-0.2143$ &  & $-0.6684$ & $-0.0016$  \\
			\hline
			$D_*$ & $-0.0628$ & $-0.3347$ & $-0.1934$ & $0.0154$ & $-2.6380$  \\
			\hline
			$z_\text{eq}$ & $0.1571$ & $0.8432$ & \multirow{6}{*}{$0$} & \multirow{3}{*}{$0$} & $0.0005$  \\
			\cline{1-3}\cline{6-6}
			$H(z_\text{eq})$ & $0.3143$ & $1.6859$ &  &  & $0.0217$  \\
			\cline{1-3}\cline{6-6}
			$k_\text{eq}$ & $0.1570$ & $0.8428$ &  &  & $0.0005$  \\
			\cline{1-3}\cline{5-6}
			$k_D$ & $0.3087$ & $0.2385$ &  & $0.4400$ & $0.0066$  \\
			\cline{1-3}\cline{5-6}
			$z_d$ & $0.0482$ & $0.0082$ &  & $0.9450$ & $0.0007$  \\
			\cline{1-3}\cline{5-6}
			$r_d$ & $-0.1687$ & $-0.2154$ &  & $-0.6162$ & $-0.0015$  \\
			\cline{1-3}\cline{4-6}
			$\sigma_8$ & $-0.1789$ & $0.6386$ & $0.2381$ & $0.0788$ & $-17.828$ \\
			\hline
			$ \mu (0.047)$ & $-0.0001$ & $-0.0006$ & $-0.0579$ & \multirow{5}{*}{$0$}  &  $-0.005$  \\
			\cline{1-4}\cline{6-6}
			$ \mu (0.142)$ & $-0.0003$ & $-0.0015$ & $-0.0517$ &  &  $-0.0132$  \\
			\cline{1-4}\cline{6-6}
			$ \mu (0.309)$ & $-0.0006$ & $-0.0031$ & $-0.0454$ &  &  $-0.0262$  \\
			\cline{1-4}\cline{6-6}
			$ \mu (1.36)$ & $-0.0016$ & $-0.0089$ & $-0.0272$ &  &  $-0.0743$  \\
			\hline
			\textbf{Ref} & $\mathbf{0.02238}$ & $\mathbf{0.1201}$ & $\mathbf{0.6732}$ & $\mathbf{v_0}$ & $\mathbf{0}$ \\
			\hline
		\end{tabular}
	}
	\caption{$Y_{|X}$ values in the axi-Higgs model. The $\Lambda$CDM(+$m_e$) model shares the values of  $Y_{|b,c,h}(+Y_{|v})$. The last row highlights the reference values of the five input parameters w.r.t. which the numbers in this table are computed.}
	\label{tab:par_derivs}
\end{table}


\section*{C. Axion Perturbations in the Axi-Higgs Model}

\begin{figure}[!ht]
	\centering
	\includegraphics[scale=0.45]{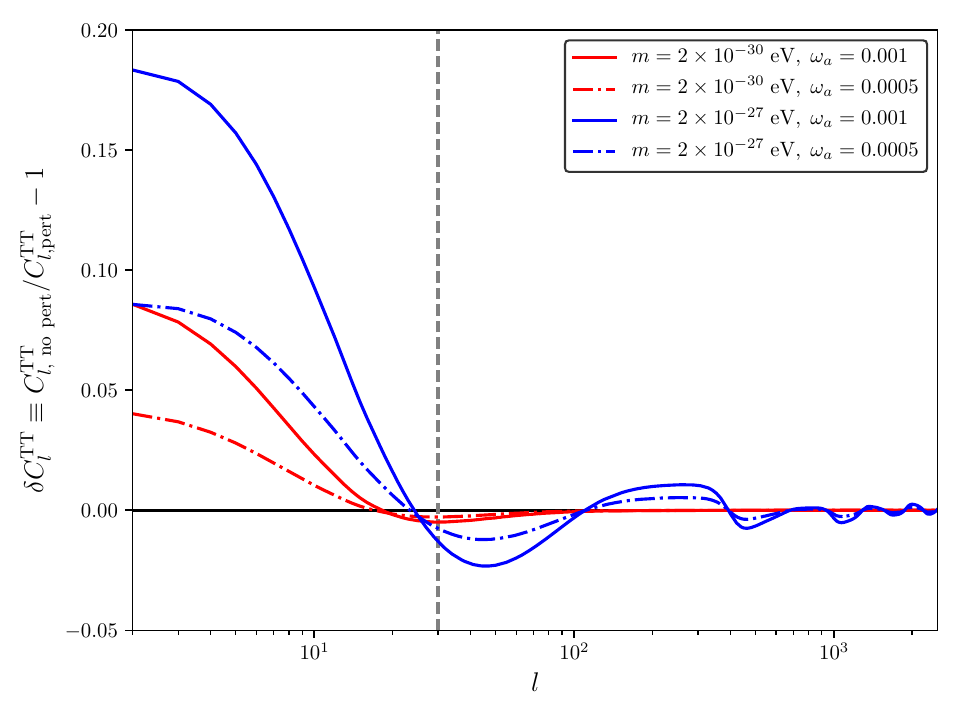}
	\includegraphics[scale=0.45]{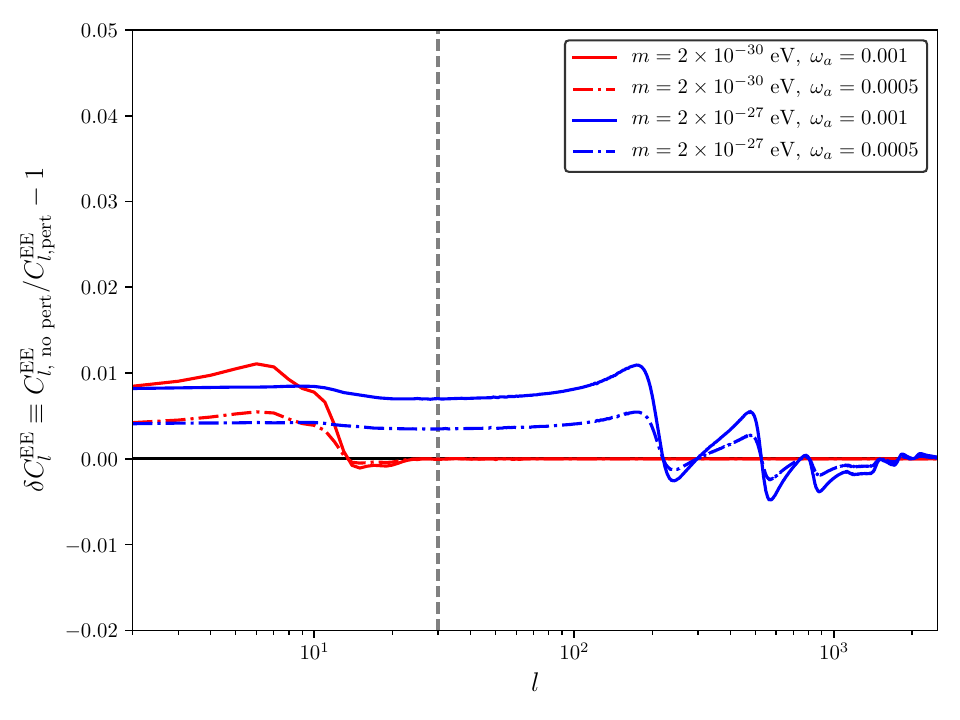}
	\caption{Residuals of the CMB TT (top) and EE (bottom) spectra, which are defined as the relative spectrum difference between the cases with and without axion perturbations. Here $m = 2 \times 10^{-30}$~eV (red; the benchmark $m$ value used in this study) and $m = 2 \times 10^{-27}$~eV (blue; a typical $m$ value for the EDE axion field) are considered. The other parameters are fixed to their best-fit values. The vertical dashed line at $l=30$ separates the low-$l$ and high-$l$ regions of the CMB power spectra. To calculate the spectra, we have applied the modified \texttt{CAMB} for the axi-Higgs model~\cite{Luu:2021yhl}.}
	\label{Fig:pert_test}
\end{figure}

The axi-Higgs and EDE models share the need of ultralight axion field, with its relic abundance $\lesssim 10^{-3}$, to mediate the Hubble flow. However, the perturbation effects caused by this axion filed are very different in magnitude. In the axi-Higgs model, $m \simeq 10^{-30} - 10^{-29}$~eV is strongly favored. The axion field rolls down from its  misaligned initial state after recombination. In contrast, the axion field in the EDE model is relatively heavy, with $m \sim 10^{-27}$~eV. The axion state transition occurs at $z\sim 10^4$~\cite{Poulin:2018cxd}. This difference yields an axion Jeans scale in the axi-Higgs model which is orders of magnitude larger than that in the EDE model. Therefore, the effects of the axion perturbation are essentially suppressed at the CMB high-$l$ region (which is characterized by a scale below the axion Jeans scale) in the axi-Higgs model, but are not for the EDE model. Actually, even in the low-$l$ region, the perturbation effects of axi-Higgs are significantly smaller than those of EDE for the TT spectrum, although they are comparable to the latter for the EE spectrum. These interesting features are shown in Fig.~\ref{Fig:pert_test}. According to this figure, the perturbation may cause a shift of $\lesssim 10\%$ and $\lesssim 1\%$ to the low-$l$ TT and EE spectra. In comparison, the uncertainties caused by cosmic variance are $> 10\%$ and $> 1\%$ respectively in these two cases (see, e.g.,~\cite{FrancoAbellan:2021hdb}). The axion perturbation effects thus can be safely neglected in the axi-Higgs model.

\section*{D. Tests of the LPA Validity} \label{app:third}

In the main text, we take a test of the LPA validity in the $\Lambda$CDM model. Here we provide more details about this test and then extend it to the $\Lambda$CDM$+m_e$ and axi-Higgs models.

In the test with $\Lambda$CDM, three model parameters and four compressed CMB observables are relevant, namely
\begin{align}
{\bf X} &= \{\omega_b, \omega_c, h\}, &\quad &N_X = 3 \ ; \\
{\bf Y} &= {\bf Y_{\rm CMB}} = \{l_a, l_\text{eq}, l_D, S_L \}, &\quad &N_Y = 4 \ .
\end{align}
The variation equations of ${\bf Y}$ are then given by 
\begin{eqnarray}
\delta l_a &=& l_{a|b} \delta\omega_b +  l_{a|c} \delta\omega_c +  l_{a|h} \delta h \nonumber \\
&=& (D_{*|b} - r_{*|b}) \delta\omega_b + (D_{*|c} - r_{*|c}) \delta\omega_c \nonumber \\ && + (D_{*|h} - r_{*|h}) \delta h \nonumber\\
&=& 0.0550\delta\omega_b - 0.1203 \delta\omega_c - 0.1934\delta h \; , \label{eq:test_1} \\ 
\delta l_\text{eq} &=& l_{\text{eq}|b} \delta\omega_b +  l_{\text{eq}|c} \delta\omega_c +  l_{\text{eq}|h} \delta h \nonumber  \\
&=& (k_{\text{eq}|b} + D_{*|b}) \delta\omega_b + (k_{\text{eq}|c} + D_{*|c}) \delta\omega_c  \nonumber \\ && + (k_{\text{eq}|h} + D_{*|h}) \delta h  \nonumber\\
&=& 0.0942\delta\omega_b + 0.5082\delta\omega_c - 0.1934\delta h \; , \label{eq:test_2} \\ 
\delta l_D &=& l_{D|b} \delta\omega_b +  l_{D|c} \delta\omega_c +  l_{D|h} \delta h \nonumber \\
&=& (k_{D|b} + D_{*|b}) \delta\omega_b + (k_{D|c} + D_{*|c}) \delta\omega_c  \nonumber \\ && + (k_{D|h} + D_{*|h}) \delta h \nonumber\\
&=& 0.2459\delta\omega_b - 0.0962\delta\omega_c - 0.1934\delta h \label{eq:test_3} \; , \\
\delta S_L &=& S_{L|b} \delta\omega_b +  S_{L|c} \delta\omega_c +  S_{L|h} \delta h \nonumber \\
&=& (\sigma_{8|b} + 0.25\Omega_{m|b}) \delta\omega_b + (\sigma_{8|c} + 0.25\Omega_{m|c}) \delta\omega_c  \nonumber \\ && + (\sigma_{8|h} + 0.25\Omega_{m|h}) \delta h \nonumber\\
&=& -0.1398\delta\omega_b + 0.8484\delta\omega_c - 0.2619\delta h \label{eq:test_4} \; ,
\end{eqnarray}
with the inputs from Tab.~\ref{tab:par_derivs_final} (or Tab.~\ref{tab:par_derivs}). The values of ${\bf Y}_{\rm o}$, ${\bf Y}_{\rm ref}$ and ${\bf \Sigma} = {\bf \Sigma}_{\rm CMB}$ can be read out from Tab.~\ref{tab:data}. Finally applying the likelihood function in Eq.~(\ref{eq:lik}) yields the results in Tab.~\ref{tab:CMBdata} and Fig.~\ref{fig:LPA}.

\begin{table}[htp]
	\resizebox{0.45\textwidth}{!}{
		\begin{tabular}{ccc}
			\hline
			\multirow{2}{*}{} & $\Lambda$CDM & $\Lambda$CDM+$m_e$ \\ 
			
			& (CMB) & (CMB+DR12) \\ 
			\hline\hline

			\multirow{2}{*}{$\omega_b$} & $0.02237\pm 0.00015$ & $0.02255\pm 0.00016$ \\
			& $0.02237 \pm 0.00014$ & $0.02255 \pm 0.00016$ \\
			\hline
			\multirow{2}{*}{$\omega_c$} & $0.1200\pm 0.0012$ & $0.1208\pm 0.0018$  \\
			& $0.1200\pm 0.0011$ & $0.1210 \pm 0.0017$  \\
			\hline					
			\multirow{2}{*}{$H_0$ (km/s/Mpc) } & $67.36\pm 0.54$ & $69.1\pm 1.2$ \\
			& $67.35 \pm 0.50$ & $69.1\pm 1.1$ \\
			\hline
			\multirow{2}{*}{$v_{\rm ini}/v_0 = m_e/m_{e,0}$} & 1 & $1.0078\pm 0.0067$ \\
			& 1 & $1.0079 \pm 0.0063$ \\
			\hline
			\multirow{2}{*}{$S_8$} & $0.832 \pm 0.013$ & $ 0.821 \pm 0.010 $ \\
			& $0.832 \pm 0.013$ & $0.823 \pm 0.010$ \\
			\hline
		\end{tabular}
	}
	\caption{Test of the LPA validity in the $\Lambda$CDM and $\Lambda$CDM+$m_e$ models. We present the results given by the full-fledged Boltzmann solver with complete spectral data of CMB combined with DR12 in the first line (which are taken from~\cite{Aghanim:2018eyx} for $\Lambda$CDM and from~\cite{Hart:2019dxi} for $\Lambda$CDM+$m_e$) and the ones obtained from the LPA in the second line in each row.}
	\label{tab:testing}
\end{table}

\begin{table}[htp]
	\resizebox{0.45\textwidth}{!}{
		\begin{tabular}{ccc}
			\hline
			\multirow{2}{*}{} & \multicolumn{2}{c}{axi-Higgs} \\ 
			
			& (CMB+BAO+WL+SN) & (CMB+BAO+WL+DR) \\ 
			\hline\hline

			\multirow{2}{*}{$\omega_b$} & $0.02269 \pm 0.00021$ & $0.02300 \pm 0.00015$ \\
			& $0.02272 \pm 0.00020$ & $0.02299 \pm 0.00014$ \\
			\hline
			\multirow{2}{*}{$\omega_c$} & $0.1205^{+0.0020}_{-0.0023}$ & $0.1233^{+0.0017}_{-0.0019}$  \\
			& $0.1205 \pm 0.0019$ & $0.1228 \pm 0.0014$  \\
			\hline					
			\multirow{2}{*}{$H_0$} & $69.5 \pm 1.6$ & $72.41 \pm 0.78$ \\
			& $69.9 \pm 1.5$ & $72.42 \pm 0.76$ \\
			\hline
			\multirow{2}{*}{$v_{\rm ini}/v_0$} & $1.0115^{+0.0087}_{-0.010}$ & $1.0272^{+0.0054}_{-0.0068}$ \\
			& $1.0123 \pm 0.0086$ & $1.0254 \pm 0.0050$ \\
			\hline
			\multirow{2}{*}{$\omega_a$} & $< 0.00205$ & $< 0.00242$ \\
			& $< 0.00125$ & $< 0.00132$ \\
			\hline
			\multirow{2}{*}{$S_8$} & $0.8061 \pm 0.0094$ & $0.7996 \pm 0.0087$ \\
			& $0.8045 \pm 0.0096$ & $0.7970 \pm 0.0088$ \\
			\hline
		\end{tabular}
	}
	\caption{Test of the LPA validity in the axi-Higgs model with $m = 2 \times 10^{-30}$~eV. The marginalized SBA and LPA posteriors for cosmological parameters are shown in the first and second lines of each row, respectively (for the LPA results in the axi-Higgs model, also see Tab.~\ref{tab:model_comparison}). Note that ``CMB'' and ``SN'' represent the set of compressed observables for the LPA while they denote the full likelihood data for the SBA.}
	\label{tab:testing_aH}
\end{table}

\begin{figure}[!ht]
	\centering
	\includegraphics[scale=0.38]{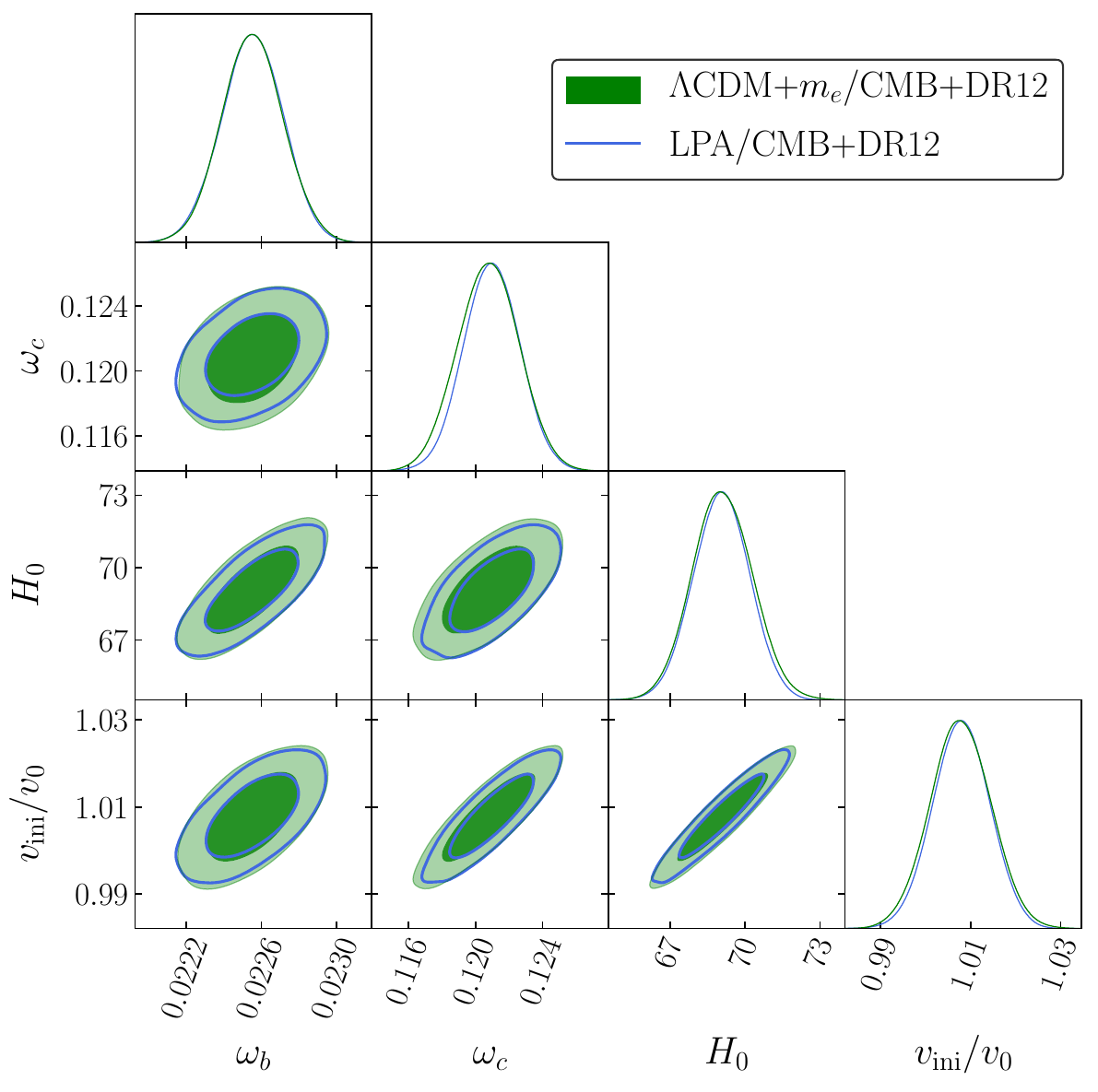}
	\caption{Comparison of the marginalized contours and posterior distributions of $\Lambda$CDM+$m_e$ between the LPA analysis and the CMB+DR12 analysis in~\cite{Hart:2019dxi}.}
	\label{fig:testing}
\end{figure}

\begin{figure*}[!ht]
	\centering
	\includegraphics[scale=0.5]{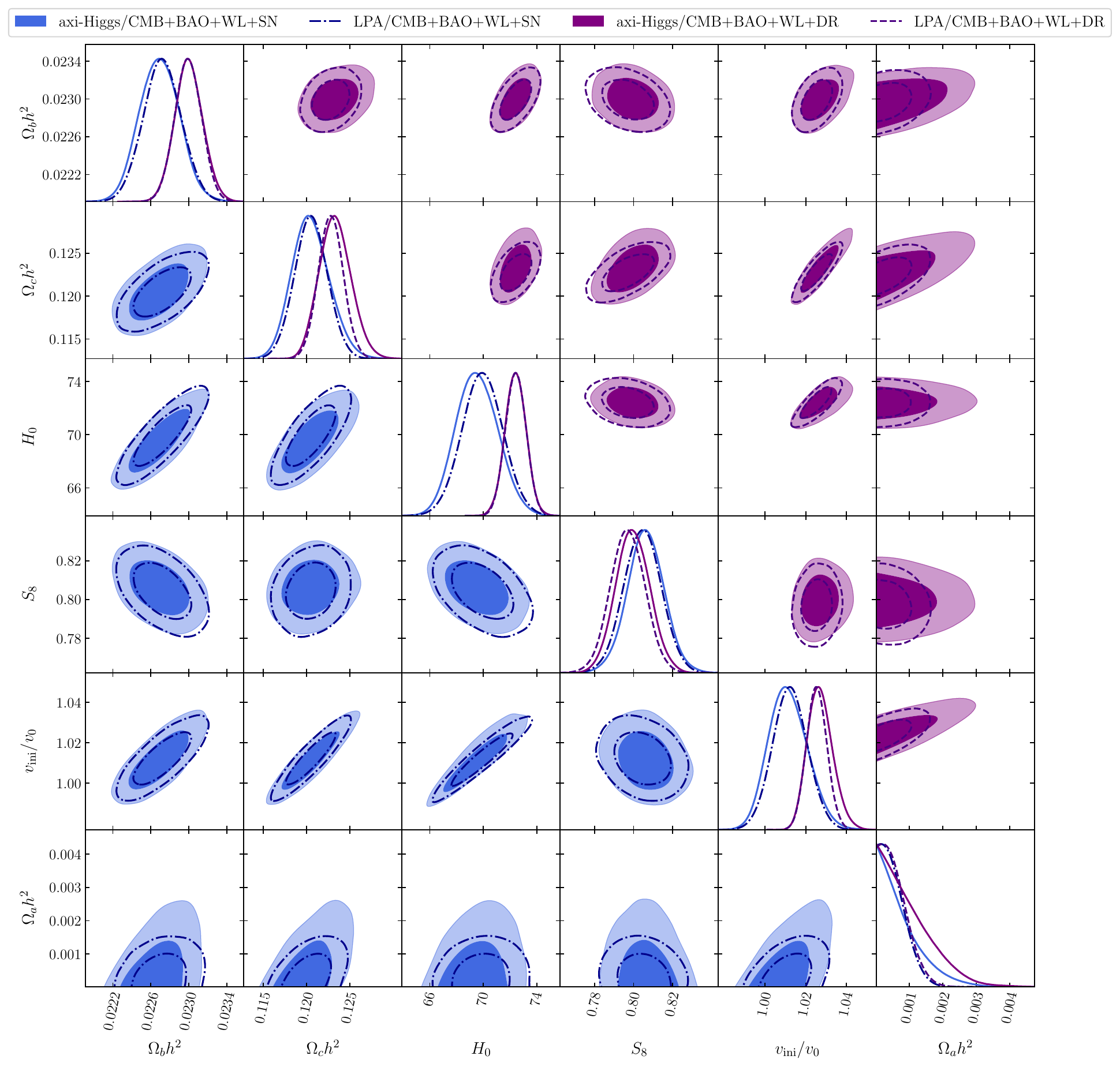}
	\caption{Comparison of the marginalized contours and posterior distributions of axi-Higgs ($m = 2 \times 10^{-30}$~eV) between the LPA (empty-dashed) and SBA (filled-solid) analyses.}
	\label{Fig:LPA_comparison}
\end{figure*}

\begin{figure}[!ht]
	\centering
	\includegraphics[scale=0.43]{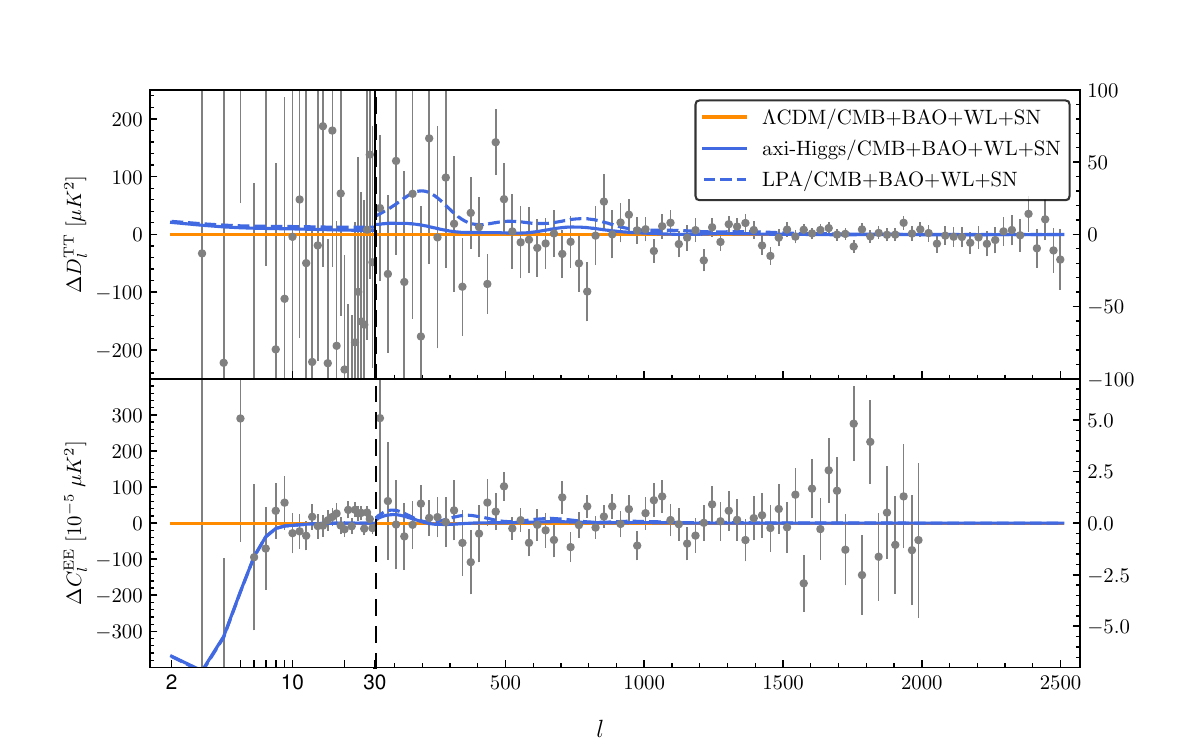}
	\includegraphics[scale=0.43]{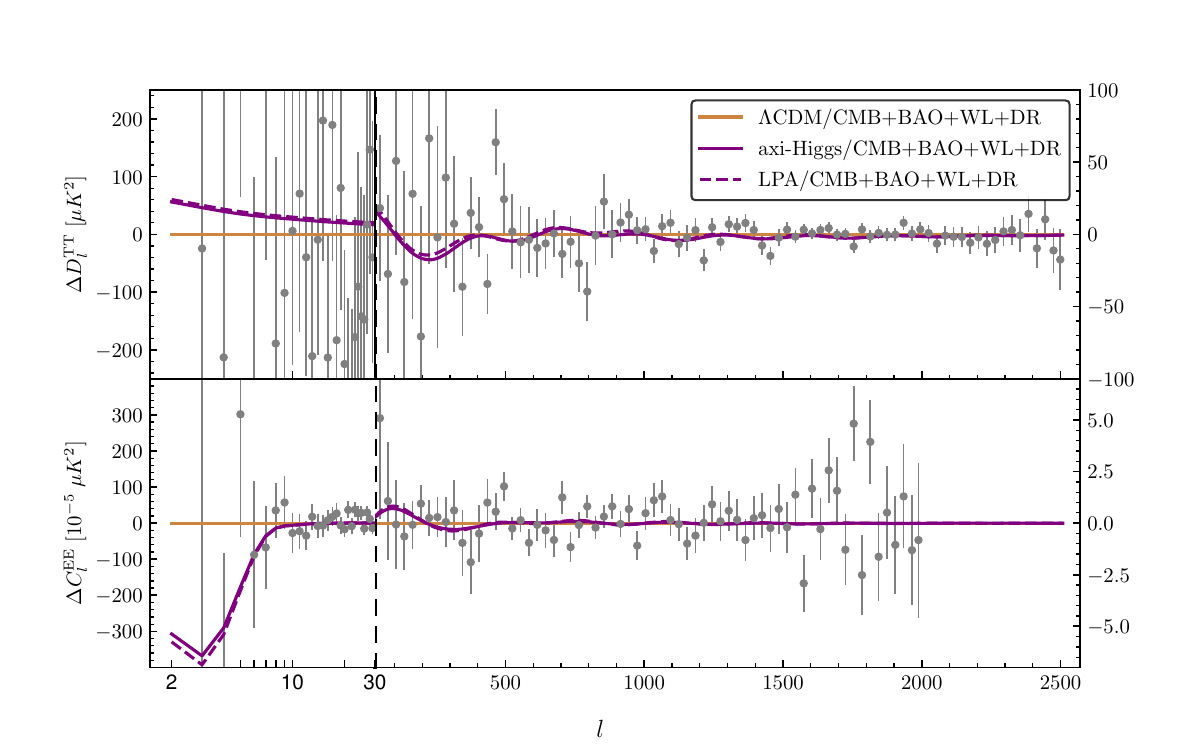}
	\caption{Residuals of the CMB spectra at the best-fit points of the axi-Higgs model w.r.t the $\Lambda$CDM model, which are fit to the CMB+BAO+WL+SN (top) and CMB+BAO+WL+DR (bottom) data. We define $D_l \equiv l(l+1)C_l / 2\pi$; $\Delta C_l^{X} \equiv C^{X}_{l, Y} - C^{X}_{l, \Lambda{\rm CDM}}$, where $X$ denotes TT/EE and $Y$ denotes LPA/SBA.}
	\label{Fig:bestfit_spectra}
\end{figure}

The test of LPA can be extended to the $\Lambda$CDM$+m_e$ model. As $\omega_a$ approaches zero, the axi-Higgs model reduces to the $\Lambda$CDM$+m_e$ model. For this purpose, we apply the CMB+DR12 data introduced in Sec.~A and Tab.~\ref{tab:data} to the relevant variation equations defined by Tab.~\ref{tab:par_derivs}. The LPA results are summarized in Tab.~\ref{tab:testing}, together with the ones given by the full-fledged Boltzmann solver (essentially a reproduction of the results in~\cite{Hart:2019dxi}), while the posterior distributions of the model parameters are presented in Fig.~\ref{fig:testing}. One can see again that the LPA analysis reproduces the results of the full-pledge Boltzmann solver nearly perfectly. In particular, this analysis yields $H_{0} = 69.1\pm 1.1$ km/s/Mpc and $v_{\rm ini}/v_0 = 1.0079 \pm 0.0063$, highly consistent with $H_{0} = 69.1\pm 1.2$ km/s/Mpc and $v_{\rm ini}/v_0 = 1.0078\pm 0.0067$ found in~\cite{Hart:2019dxi}, and hence ``re-discovers'' that varying $m_e$ can result in a bigger $H_0$ value than the $\Lambda$CDM one which is pointed out in~\cite{Ade:2014zfo,Hart:2019dxi}.

\begin{figure}[!ht]
	\centering
	\includegraphics[scale=0.33]{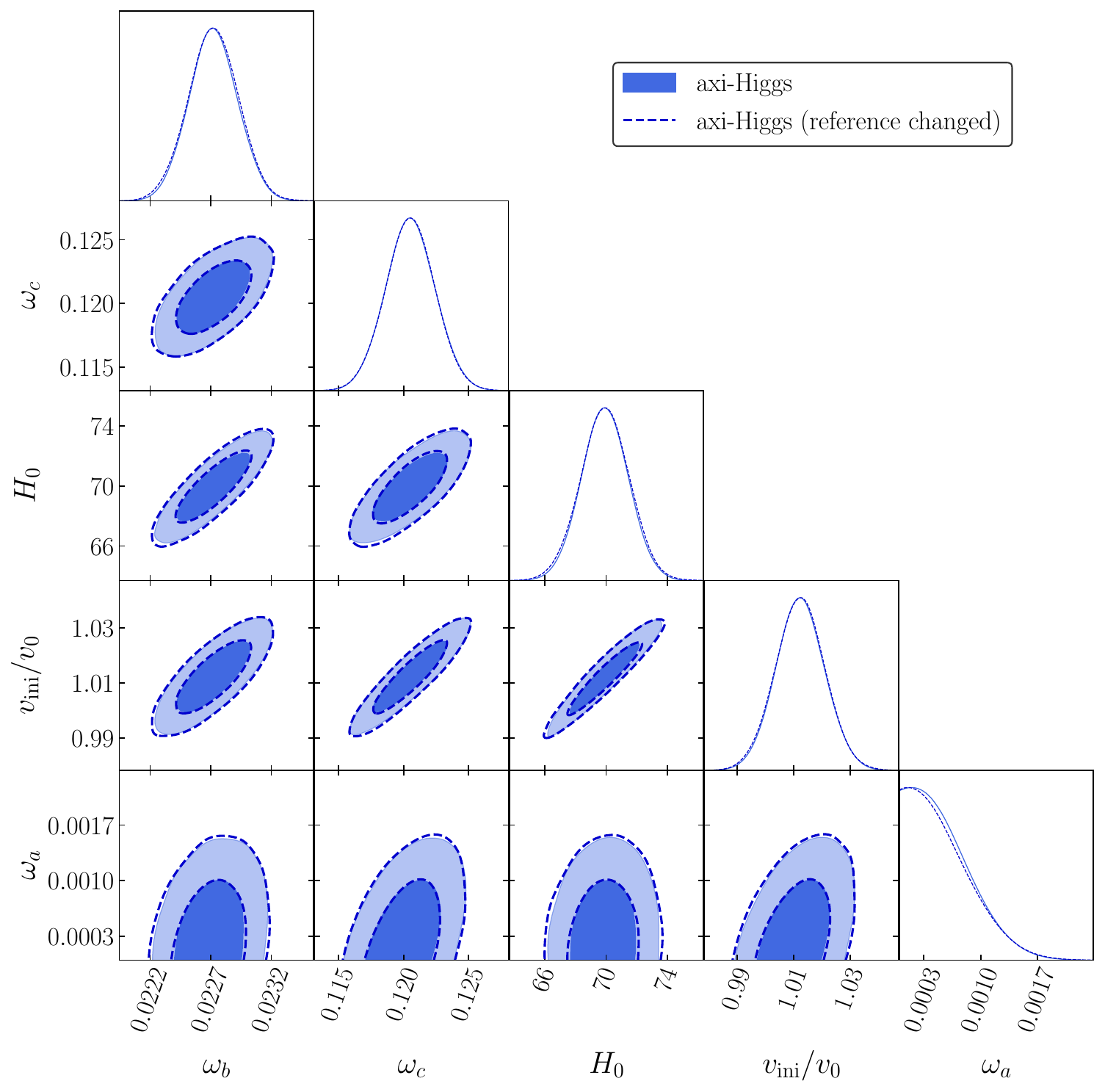}
	\caption{Comparison of the LPA contours in the axi-Higgs model with different reference points. Here the new reference point (dashed) is defined as $(\omega_b, \omega_c, h, v_{\rm ini}, \omega_a)=(0.02276, 0.1209, 0.703, 1.0142v_0, 0)$. }
	\label{fig:LPArho}
\end{figure}

The recently developed codes for standard Boltzmann analysis (SBA) in the axi-Higgs model~\cite{Luu:2021yhl} allow us to extend the LPA test to this model also. For this purpose, we present both LPA and SBA posteriors for cosmological parameters in Tab.~\ref{tab:testing_aH} and Fig.~\ref{Fig:LPA_comparison}. We also show the residuals of the CMB spectra at the LPA and SBA best-fit points in Fig.~\ref{Fig:bestfit_spectra}. By taking a comparison, one can see that the LPA and SBA results agree with each other reasonably well, which further confirms the LPA validity.

At last, we test the impacts of the choice of reference point on the LPA results. For this purpose, we artificially define a new reference point which is away  from the original one by several percents in $\omega_b$, $\omega_c$, $h$ and $v_{\rm ini}$, namely $\{ \omega_b, \omega_c, h, v_{\rm ini}, \omega_a \} = \{ 0.02276, 0.1209, 0.703, 1.0142 v_0, 0 \}$. The LPA contours in the axi-Higgs model with these two reference points are shown in Fig.~\ref{fig:LPArho}. The negligible discrepancy between these contours indicates that the LPA is not very sensitive to the choice of the reference point, as long as the new choice is not globally far.

\section*{E. Impacts of the Ultralight-Axion Mass} \label{app:fourth}

We consider the impacts of the axion mass $m$, by varying its value from $10^{-28}$~eV to $10^{-31}$~eV. For $m = 10^{-28}$\,eV, we have $z_a\sim 2100$ and $z_{\rm eq} > z_a > z_*$. The impact of $a$ on $l_{\rm eq}$ is still given by Eq.~(\ref{eq:dleq}) approximately, but it starts to roll down before $z_*$. So keeping $\delta v_{\rm rec}$ at a few percent level requires a larger $C$ value. For $m =10^{-27}\,$eV, $a$ contributes to DM at $z_*$. At the smaller $m$ end, for $m =10^{-31}\,$eV, $a$ behaves like DE as its dropping is late, which results in a weakened suppression for galactic clustering or $S_8$. Notably, such a small $m$ has been ruled out by the AC measurements~\cite{Fung:2021wbz}.

\begin{figure}[!ht]
	\centering
	\includegraphics[scale=0.4]{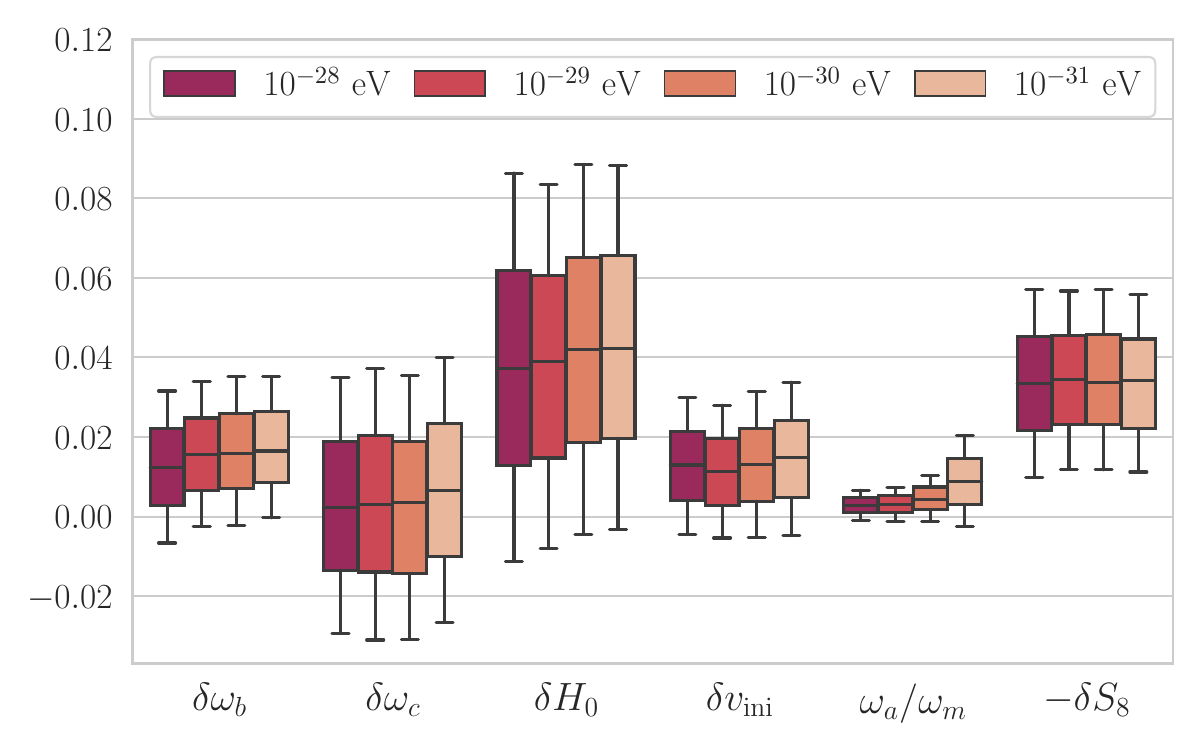}
	\caption{Relative variations of the marginalized parameter values as a function of the axion mass $m$, in the axi-Higgs/CMB+BAO+WL +SN analysis scenario. The central boxes and error bars denote the uncertainties at 68\% and 95\% confidence level.}
	\label{fig:axion_mass}
\end{figure}

We demonstrate the relative variations of the marginalized parameter values in the axi-Higgs/CMB+BAO+WL +SN analysis scenario as a function of axion mass in Fig.~\ref{fig:axion_mass}. We also present the marginalized constraints on the axi-Higgs model with different axion masses in Tab.\,\ref{tab:axion_mass}.  In the bottom subtable, the SN data have been replaced with the DR measurements.

\begin{table}[htp]
	\resizebox{0.49\textwidth}{!}{
		\begin{tabular}{|c|c|c|c|c|}
			\hline
			Model & \multicolumn{4}{c|}{axi-Higgs/CMB+BAO+WL+SN} \\
			\hline
			$m$ & \multicolumn{1}{c}{$10^{-28}$ eV} & \multicolumn{1}{c}{$10^{-29}$ eV} & \multicolumn{1}{c}{$10^{-30}$ eV} & \multicolumn{1}{c|}{$10^{-31}$ eV} \\ 
			\hline\hline
			$\omega_b$ & $0.02264 \pm 0.00020$ & $0.02271 \pm 0.00020$ & $0.02273 \pm 0.00020$ & $0.02278 \pm 0.00021$ \\
			
			$\omega_c$ & $0.1205 \pm 0.0019$ & $0.1204 \pm 0.0019$ & $0.1205 \pm 0.0019$ & $0.1207 \pm 0.0019$ \\
			
			$H_0$ & $69.8 \pm 1.5$ & $69.9 \pm 1.5$ & $69.9 \pm 1.5$ & $70.0 \pm 1.5$ \\
			
			$v_{\rm ini}/v_0$ & $1.0113 \pm 0.0085$ & $1.0115 \pm 0.0085$ & $1.0126 \pm 0.0087$ & $1.0143 \pm 0.0092$ \\
			
			$\omega_a$ & $<0.000942$ & $<0.000978$ & $< 0.00146$ & $< 0.00275$ \\
			
			\hline
			
			$\omega_m$ & $0.1442 \pm 0.0021$ & $0.1442 \pm 0.0021$ & $0.1445 \pm 0.0022$ & $0.1453^{+0.0023}_{-0.0028}$ \\
			
			$S_8$ & $0.8046 \pm 0.0095$ & $0.8046 \pm 0.0094$ & $0.8044 \pm 0.0096$ & $0.8051 \pm 0.0095$ \\
			\hline
		\end{tabular}
	}

	\bigskip
	
	\resizebox{0.49\textwidth}{!}{
		\begin{tabular}{|c|c|c|c|c|}
			\hline
			Model & \multicolumn{4}{c|}{axi-Higgs/CMB+BAO+WL+DR} \\ 
			\hline
			$m$ & \multicolumn{1}{c}{$10^{-28}$ eV} & \multicolumn{1}{c}{$10^{-29}$ eV} & \multicolumn{1}{c}{$10^{-30}$ eV} & \multicolumn{1}{c|}{$10^{-31}$ eV} \\ 
			\hline\hline
			$\omega_b$ & $0.02288 \pm 0.00014$ & $0.02297 \pm 0.00014$ & $0.02300 \pm 0.00014$ & $0.02307 \pm 0.00016$ \\
			
			$\omega_c$ & $0.1227 \pm 0.0014$ & $0.1228 \pm 0.0014$ & $0.1229 \pm 0.0015$ & $0.1231 \pm 0.0015$ \\
			
			$H_0$ & $72.39 \pm 0.76$ & $72.41 \pm 0.76$ & $72.43 \pm 0.77$ & $72.46 \pm 0.76$ \\
			
			$v_{\rm ini}/v_0$ & $1.0242 \pm 0.0047$ & $1.0249 \pm 0.0049$ & $1.0258 \pm 0.0052$ & $1.0283^{+0.0055}_{-0.0066}$ \\
			
			$\omega_a$ & $< 0.000925$ & $< 0.00102$ & $< 0.00152$ & $< 0.00324$ \\
			
			\hline
			
			$\omega_m$ & $0.1466 \pm 0.0016$ & $0.1468 \pm 0.0016$ & $0.1472^{+0.0016}_{-0.0018}$ & $0.1482^{+0.0019}_{-0.0025}$ \\
			
			$S_8$ & $0.7964^{+0.0093}_{-0.0084}
			$ & $0.7968 \pm 0.0088$ & $0.7970 \pm 0.0088$ & $0.7977 \pm 0.0087$ \\
			\hline
		\end{tabular}
	}
	
	\caption{Marginalized parameter values in the axi-Higgs model with different axion mass. }
	\label{tab:axion_mass}
\end{table}

\section*{F. More on Axi-Higgs Cosmology} \label{app:fifth}
\begin{figure}[!ht]
	\centering
	\includegraphics[scale=0.54]{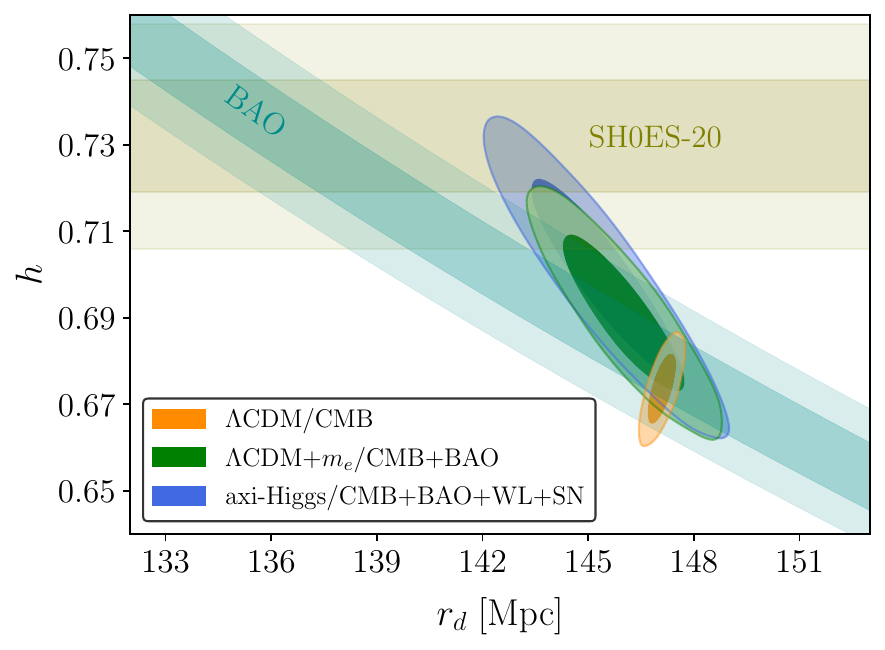}
	\caption{Cosmological constraints on $r_d$ and $h$ in different models. The local measurement of $H_0 = 73.2 \pm 1.3$~km/s/Mpc is taken from SH0ES-20~\cite{Riess:2020fzl} using distance ladder (the horizontal olive bands), while the combined BAO constraint of $r_d h = 99.95 \pm 1.20$~Mpc (the shaded cyan bands) is taken from~\cite{Pogosian:2020ded}.}	 
	\label{fig:rdh}
\end{figure}

In~\cite{Fung:2021wbz}, we give a crude estimate of $\delta v_{\rm rec}$ and $\omega_c$ in the axi-Higgs model, which are more precisely determined here.
We also explore the correlation between $h$ and $r_d$, following~\cite{Pogosian:2020ded}, and demonstrate the relevant results in its Fig.~3. Below we update that figure to Fig.~\ref{fig:rdh} here. Different from its original version~\cite{Fung:2021wbz}, where $\delta v_{\rm rec}$ and $\omega_b$ are set as inputs of the benchmarks using, e.g., the favored values by the BBN data, we allow these two parameters to vary freely to fit data in this study. This explains the difference between these two figures.

\begin{figure}[!ht]
	\centering
	\includegraphics[scale=0.54]{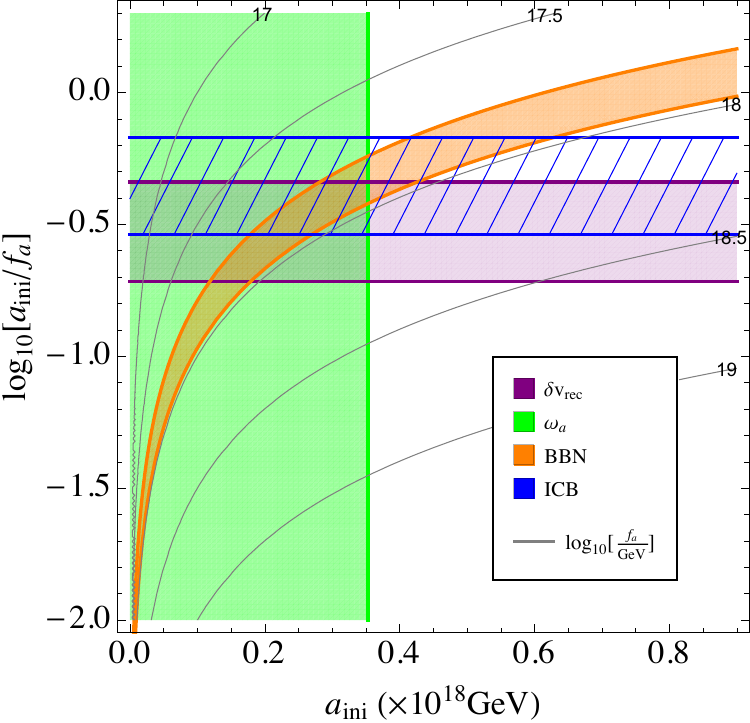}
	\caption{Overall picture on the axi-Higgs cosmology.}
	\label{fig:summary}
\end{figure}

In Fig.~\ref{fig:summary}, we present an overall picture on the axi-Higgs cosmology in the benchmark scenario. With $C' = C {f_a}^2/2M_{\rm Pl}^2 = 0.1$, we recast the marginalized CMB+BAO+WL+SN limits for $\omega_a$ and $\delta v_{\rm rec} (= \delta v_{\rm ini})$ as the constraints for $a_{\rm ini}$ and $\frac{a_{\rm ini}}{f_a}$ (the green and purple shaded bands). The $H_0$ and $S_8$ tensions can be relaxed in this context. The blue hatch region encodes the observed ICB in the polarized CMB data~\cite{Minami:2020odp}, with $C_\gamma = 2$~\cite{Fung:2021wbz}. The $\delta v_{\rm BBN}$ values favored to address the $^7$Li puzzle~\cite{Fung:2021wbz}, with $\omega_a$ from the CMB+BAO+WL+SN data as an input, is projected to the constraint for $f_a$ (the orange shaded region) through Eq.~(\ref{model1}). In the intersection region where all of the four puzzles are addressed simultaneously, $f_a$ is favored to be $\sim 10^{17} - 10^{18}\,$~GeV. 

In the brane-world scenario of string theory, the axi-Higgs axion is expected to come from the closed string sector, where the graviton also comes from. This axion is a component of the complex-structure moduli responsible for the shape of Calabi-Yau orientifold or the dilaton mode. This is different from an axion in  the open string sector inside a (anti-D3-)brane stack, in which all standard model particles reside. So, a decay constant $f_a$ lying somewhere between the GUT-string scale and $M_{\rm Pl}$, as it does, is very natural!

\bigskip{}

\bibliography{reference.bib}

\begin{thebibliography}{68}
\expandafter\ifx\csname natexlab\endcsname\relax\def\natexlab#1{#1}\fi
\expandafter\ifx\csname bibnamefont\endcsname\relax
  \def\bibnamefont#1{#1}\fi
\expandafter\ifx\csname bibfnamefont\endcsname\relax
  \def\bibfnamefont#1{#1}\fi
\expandafter\ifx\csname citenamefont\endcsname\relax
  \def\citenamefont#1{#1}\fi
\expandafter\ifx\csname url\endcsname\relax
  \def\url#1{\texttt{#1}}\fi
\expandafter\ifx\csname urlprefix\endcsname\relax\def\urlprefix{URL }\fi
\providecommand{\bibinfo}[2]{#2}
\providecommand{\eprint}[2][]{\url{#2}}

\bibitem[{\citenamefont{Aghanim et~al.}(2020{\natexlab{a}})}]{Aghanim:2018eyx}
\bibinfo{author}{\bibfnamefont{N.}~\bibnamefont{Aghanim}} \bibnamefont{et~al.}
  (\bibinfo{collaboration}{Planck}), \bibinfo{journal}{Astron. Astrophys.}
  \textbf{\bibinfo{volume}{641}}, \bibinfo{pages}{A6}
  (\bibinfo{year}{2020}{\natexlab{a}}), \eprint{1807.06209}.

\bibitem[{\citenamefont{Verde et~al.}(2019)\citenamefont{Verde, Treu, and
  Riess}}]{Verde:2019ivm}
\bibinfo{author}{\bibfnamefont{L.}~\bibnamefont{Verde}},
  \bibinfo{author}{\bibfnamefont{T.}~\bibnamefont{Treu}}, \bibnamefont{and}
  \bibinfo{author}{\bibfnamefont{A.}~\bibnamefont{Riess}},
  \bibinfo{journal}{Nature Astron.} \textbf{\bibinfo{volume}{3}},
  \bibinfo{pages}{891} (\bibinfo{year}{2019}), \eprint{1907.10625}.

\bibitem[{\citenamefont{Knox and Millea}(2020)}]{Knox:2019rjx}
\bibinfo{author}{\bibfnamefont{L.}~\bibnamefont{Knox}} \bibnamefont{and}
  \bibinfo{author}{\bibfnamefont{M.}~\bibnamefont{Millea}},
  \bibinfo{journal}{Phys. Rev. D} \textbf{\bibinfo{volume}{101}},
  \bibinfo{pages}{043533} (\bibinfo{year}{2020}), \eprint{1908.03663}.

\bibitem[{\citenamefont{Di~Valentino et~al.}(2021)\citenamefont{Di~Valentino,
  Mena, Pan, Visinelli, Yang, Melchiorri, Mota, Riess, and
  Silk}}]{DiValentino:2021izs}
\bibinfo{author}{\bibfnamefont{E.}~\bibnamefont{Di~Valentino}},
  \bibinfo{author}{\bibfnamefont{O.}~\bibnamefont{Mena}},
  \bibinfo{author}{\bibfnamefont{S.}~\bibnamefont{Pan}},
  \bibinfo{author}{\bibfnamefont{L.}~\bibnamefont{Visinelli}},
  \bibinfo{author}{\bibfnamefont{W.}~\bibnamefont{Yang}},
  \bibinfo{author}{\bibfnamefont{A.}~\bibnamefont{Melchiorri}},
  \bibinfo{author}{\bibfnamefont{D.~F.} \bibnamefont{Mota}},
  \bibinfo{author}{\bibfnamefont{A.~G.} \bibnamefont{Riess}}, \bibnamefont{and}
  \bibinfo{author}{\bibfnamefont{J.}~\bibnamefont{Silk}}
  (\bibinfo{year}{2021}), \eprint{2103.01183}.

\bibitem[{\citenamefont{Heymans et~al.}(2021)}]{Heymans:2020gsg}
\bibinfo{author}{\bibfnamefont{C.}~\bibnamefont{Heymans}} \bibnamefont{et~al.},
  \bibinfo{journal}{Astron. Astrophys.} \textbf{\bibinfo{volume}{646}},
  \bibinfo{pages}{A140} (\bibinfo{year}{2021}), \eprint{2007.15632}.

\bibitem[{\citenamefont{Abbott et~al.}(2018)}]{Abbott:2017wau}
\bibinfo{author}{\bibfnamefont{T.~M.~C.} \bibnamefont{Abbott}}
  \bibnamefont{et~al.} (\bibinfo{collaboration}{DES}), \bibinfo{journal}{Phys.
  Rev. D} \textbf{\bibinfo{volume}{98}}, \bibinfo{pages}{043526}
  (\bibinfo{year}{2018}), \eprint{1708.01530}.

\bibitem[{\citenamefont{Fung et~al.}(2021)\citenamefont{Fung, Li, Liu, Luu,
  Qiu, and Tye}}]{Fung:2021wbz}
\bibinfo{author}{\bibfnamefont{L.~W.~H.} \bibnamefont{Fung}},
  \bibinfo{author}{\bibfnamefont{L.}~\bibnamefont{Li}},
  \bibinfo{author}{\bibfnamefont{T.}~\bibnamefont{Liu}},
  \bibinfo{author}{\bibfnamefont{H.~N.} \bibnamefont{Luu}},
  \bibinfo{author}{\bibfnamefont{Y.-C.} \bibnamefont{Qiu}}, \bibnamefont{and}
  \bibinfo{author}{\bibfnamefont{S.~H.~H.} \bibnamefont{Tye}},
  \bibinfo{journal}{JCAP} \textbf{\bibinfo{volume}{08}}, \bibinfo{pages}{057}
  (\bibinfo{year}{2021}), \eprint{2102.11257}.

\bibitem[{\citenamefont{Kneller and McLaughlin}(2003)}]{Kneller:2003xf}
\bibinfo{author}{\bibfnamefont{J.~P.} \bibnamefont{Kneller}} \bibnamefont{and}
  \bibinfo{author}{\bibfnamefont{G.~C.} \bibnamefont{McLaughlin}},
  \bibinfo{journal}{Phys. Rev. D} \textbf{\bibinfo{volume}{68}},
  \bibinfo{pages}{103508} (\bibinfo{year}{2003}), \eprint{nucl-th/0305017}.

\bibitem[{\citenamefont{Li and Chu}(2006)}]{Li:2005km}
\bibinfo{author}{\bibfnamefont{B.}~\bibnamefont{Li}} \bibnamefont{and}
  \bibinfo{author}{\bibfnamefont{M.-C.} \bibnamefont{Chu}},
  \bibinfo{journal}{Phys. Rev. D} \textbf{\bibinfo{volume}{73}},
  \bibinfo{pages}{023509} (\bibinfo{year}{2006}), \eprint{astro-ph/0511642}.

\bibitem[{\citenamefont{Coc et~al.}(2007)\citenamefont{Coc, Nunes, Olive, Uzan,
  and Vangioni}}]{Coc:2006sx}
\bibinfo{author}{\bibfnamefont{A.}~\bibnamefont{Coc}},
  \bibinfo{author}{\bibfnamefont{N.~J.} \bibnamefont{Nunes}},
  \bibinfo{author}{\bibfnamefont{K.~A.} \bibnamefont{Olive}},
  \bibinfo{author}{\bibfnamefont{J.-P.} \bibnamefont{Uzan}}, \bibnamefont{and}
  \bibinfo{author}{\bibfnamefont{E.}~\bibnamefont{Vangioni}},
  \bibinfo{journal}{Phys. Rev. D} \textbf{\bibinfo{volume}{76}},
  \bibinfo{pages}{023511} (\bibinfo{year}{2007}), \eprint{astro-ph/0610733}.

\bibitem[{\citenamefont{Dent et~al.}(2007)\citenamefont{Dent, Stern, and
  Wetterich}}]{Dent:2007zu}
\bibinfo{author}{\bibfnamefont{T.}~\bibnamefont{Dent}},
  \bibinfo{author}{\bibfnamefont{S.}~\bibnamefont{Stern}}, \bibnamefont{and}
  \bibinfo{author}{\bibfnamefont{C.}~\bibnamefont{Wetterich}},
  \bibinfo{journal}{Phys. Rev. D} \textbf{\bibinfo{volume}{76}},
  \bibinfo{pages}{063513} (\bibinfo{year}{2007}), \eprint{0705.0696}.

\bibitem[{\citenamefont{Browder et~al.}(2009)\citenamefont{Browder, Gershon,
  Pirjol, Soni, and Zupan}}]{Browder:2008em}
\bibinfo{author}{\bibfnamefont{T.~E.} \bibnamefont{Browder}},
  \bibinfo{author}{\bibfnamefont{T.}~\bibnamefont{Gershon}},
  \bibinfo{author}{\bibfnamefont{D.}~\bibnamefont{Pirjol}},
  \bibinfo{author}{\bibfnamefont{A.}~\bibnamefont{Soni}}, \bibnamefont{and}
  \bibinfo{author}{\bibfnamefont{J.}~\bibnamefont{Zupan}},
  \bibinfo{journal}{Rev. Mod. Phys.} \textbf{\bibinfo{volume}{81}},
  \bibinfo{pages}{1887} (\bibinfo{year}{2009}), \eprint{0802.3201}.

\bibitem[{\citenamefont{Bedaque et~al.}(2011)\citenamefont{Bedaque, Luu, and
  Platter}}]{Bedaque:2010hr}
\bibinfo{author}{\bibfnamefont{P.~F.} \bibnamefont{Bedaque}},
  \bibinfo{author}{\bibfnamefont{T.}~\bibnamefont{Luu}}, \bibnamefont{and}
  \bibinfo{author}{\bibfnamefont{L.}~\bibnamefont{Platter}},
  \bibinfo{journal}{Phys. Rev. C} \textbf{\bibinfo{volume}{83}},
  \bibinfo{pages}{045803} (\bibinfo{year}{2011}), \eprint{1012.3840}.

\bibitem[{\citenamefont{Cheoun et~al.}(2011)\citenamefont{Cheoun, Kajino,
  Kusakabe, and Mathews}}]{Cheoun:2011yn}
\bibinfo{author}{\bibfnamefont{M.-K.} \bibnamefont{Cheoun}},
  \bibinfo{author}{\bibfnamefont{T.}~\bibnamefont{Kajino}},
  \bibinfo{author}{\bibfnamefont{M.}~\bibnamefont{Kusakabe}}, \bibnamefont{and}
  \bibinfo{author}{\bibfnamefont{G.~J.} \bibnamefont{Mathews}},
  \bibinfo{journal}{Phys. Rev. D} \textbf{\bibinfo{volume}{84}},
  \bibinfo{pages}{043001} (\bibinfo{year}{2011}), \eprint{1104.5547}.

\bibitem[{\citenamefont{Berengut et~al.}(2013)\citenamefont{Berengut, Epelbaum,
  Flambaum, Hanhart, Meissner, Nebreda, and Pelaez}}]{Berengut:2013nh}
\bibinfo{author}{\bibfnamefont{J.}~\bibnamefont{Berengut}},
  \bibinfo{author}{\bibfnamefont{E.}~\bibnamefont{Epelbaum}},
  \bibinfo{author}{\bibfnamefont{V.}~\bibnamefont{Flambaum}},
  \bibinfo{author}{\bibfnamefont{C.}~\bibnamefont{Hanhart}},
  \bibinfo{author}{\bibfnamefont{U.-G.} \bibnamefont{Meissner}},
  \bibinfo{author}{\bibfnamefont{J.}~\bibnamefont{Nebreda}}, \bibnamefont{and}
  \bibinfo{author}{\bibfnamefont{J.}~\bibnamefont{Pelaez}},
  \bibinfo{journal}{Phys. Rev. D} \textbf{\bibinfo{volume}{87}},
  \bibinfo{pages}{085018} (\bibinfo{year}{2013}), \eprint{1301.1738}.

\bibitem[{\citenamefont{Hall et~al.}(2014)\citenamefont{Hall, Pinner, and
  Ruderman}}]{Hall:2014dfa}
\bibinfo{author}{\bibfnamefont{L.~J.} \bibnamefont{Hall}},
  \bibinfo{author}{\bibfnamefont{D.}~\bibnamefont{Pinner}}, \bibnamefont{and}
  \bibinfo{author}{\bibfnamefont{J.~T.} \bibnamefont{Ruderman}},
  \bibinfo{journal}{JHEP} \textbf{\bibinfo{volume}{12}}, \bibinfo{pages}{134}
  (\bibinfo{year}{2014}), \eprint{1409.0551}.

\bibitem[{\citenamefont{Heffernan et~al.}(2017)\citenamefont{Heffernan,
  Banerjee, and Walker-Loud}}]{Heffernan:2017hwa}
\bibinfo{author}{\bibfnamefont{M.}~\bibnamefont{Heffernan}},
  \bibinfo{author}{\bibfnamefont{P.}~\bibnamefont{Banerjee}}, \bibnamefont{and}
  \bibinfo{author}{\bibfnamefont{A.}~\bibnamefont{Walker-Loud}}
  (\bibinfo{year}{2017}), \eprint{1706.04991}.

\bibitem[{\citenamefont{Mori and Kusakabe}(2019)}]{Mori:2019cfo}
\bibinfo{author}{\bibfnamefont{K.}~\bibnamefont{Mori}} \bibnamefont{and}
  \bibinfo{author}{\bibfnamefont{M.}~\bibnamefont{Kusakabe}},
  \bibinfo{journal}{Phys. Rev. D} \textbf{\bibinfo{volume}{99}},
  \bibinfo{pages}{083013} (\bibinfo{year}{2019}), \eprint{1901.03943}.

\bibitem[{\citenamefont{Minami and Komatsu}(2020)}]{Minami:2020odp}
\bibinfo{author}{\bibfnamefont{Y.}~\bibnamefont{Minami}} \bibnamefont{and}
  \bibinfo{author}{\bibfnamefont{E.}~\bibnamefont{Komatsu}},
  \bibinfo{journal}{Phys. Rev. Lett.} \textbf{\bibinfo{volume}{125}},
  \bibinfo{pages}{221301} (\bibinfo{year}{2020}), \eprint{2011.11254}.

\bibitem[{\citenamefont{Jedamzik et~al.}(2021)\citenamefont{Jedamzik, Pogosian,
  and Zhao}}]{Jedamzik:2020zmd}
\bibinfo{author}{\bibfnamefont{K.}~\bibnamefont{Jedamzik}},
  \bibinfo{author}{\bibfnamefont{L.}~\bibnamefont{Pogosian}}, \bibnamefont{and}
  \bibinfo{author}{\bibfnamefont{G.-B.} \bibnamefont{Zhao}},
  \bibinfo{journal}{Commun. in Phys.} \textbf{\bibinfo{volume}{4}},
  \bibinfo{pages}{123} (\bibinfo{year}{2021}), \eprint{2010.04158}.

\bibitem[{\citenamefont{Sekiguchi and Takahashi}(2021)}]{Sekiguchi:2020teg}
\bibinfo{author}{\bibfnamefont{T.}~\bibnamefont{Sekiguchi}} \bibnamefont{and}
  \bibinfo{author}{\bibfnamefont{T.}~\bibnamefont{Takahashi}},
  \bibinfo{journal}{Phys. Rev. D} \textbf{\bibinfo{volume}{103}},
  \bibinfo{pages}{083507} (\bibinfo{year}{2021}), \eprint{2007.03381}.

\bibitem[{\citenamefont{Marsh}(2016)}]{Marsh:2015xka}
\bibinfo{author}{\bibfnamefont{D.~J.~E.} \bibnamefont{Marsh}},
  \bibinfo{journal}{Phys. Rept.} \textbf{\bibinfo{volume}{643}},
  \bibinfo{pages}{1} (\bibinfo{year}{2016}), \eprint{1510.07633}.

\bibitem[{\citenamefont{Hlozek et~al.}(2015)\citenamefont{Hlozek, Grin, Marsh,
  and Ferreira}}]{Hlozek:2014lca}
\bibinfo{author}{\bibfnamefont{R.}~\bibnamefont{Hlozek}},
  \bibinfo{author}{\bibfnamefont{D.}~\bibnamefont{Grin}},
  \bibinfo{author}{\bibfnamefont{D.~J.~E.} \bibnamefont{Marsh}},
  \bibnamefont{and} \bibinfo{author}{\bibfnamefont{P.~G.}
  \bibnamefont{Ferreira}}, \bibinfo{journal}{Phys. Rev. D}
  \textbf{\bibinfo{volume}{91}}, \bibinfo{pages}{103512}
  (\bibinfo{year}{2015}), \eprint{1410.2896}.

\bibitem[{\citenamefont{Hlozek et~al.}(2018)\citenamefont{Hlozek, Marsh, and
  Grin}}]{Hlozek:2017zzf}
\bibinfo{author}{\bibfnamefont{R.}~\bibnamefont{Hlozek}},
  \bibinfo{author}{\bibfnamefont{D.~J.~E.} \bibnamefont{Marsh}},
  \bibnamefont{and} \bibinfo{author}{\bibfnamefont{D.}~\bibnamefont{Grin}},
  \bibinfo{journal}{Mon. Not. Roy. Astron. Soc.}
  \textbf{\bibinfo{volume}{476}}, \bibinfo{pages}{3063} (\bibinfo{year}{2018}),
  \eprint{1708.05681}.

\bibitem[{\citenamefont{Franco~Abell\'an
  et~al.}(2022)\citenamefont{Franco~Abell\'an, Chacko, Dev, Du, Poulin, and
  Tsai}}]{FrancoAbellan:2021hdb}
\bibinfo{author}{\bibfnamefont{G.}~\bibnamefont{Franco~Abell\'an}},
  \bibinfo{author}{\bibfnamefont{Z.}~\bibnamefont{Chacko}},
  \bibinfo{author}{\bibfnamefont{A.}~\bibnamefont{Dev}},
  \bibinfo{author}{\bibfnamefont{P.}~\bibnamefont{Du}},
  \bibinfo{author}{\bibfnamefont{V.}~\bibnamefont{Poulin}}, \bibnamefont{and}
  \bibinfo{author}{\bibfnamefont{Y.}~\bibnamefont{Tsai}},
  \bibinfo{journal}{JHEP} \textbf{\bibinfo{volume}{08}}, \bibinfo{pages}{076}
  (\bibinfo{year}{2022}), \eprint{2112.13862}.

\bibitem[{\citenamefont{Karwal and Kamionkowski}(2016)}]{Karwal:2016vyq}
\bibinfo{author}{\bibfnamefont{T.}~\bibnamefont{Karwal}} \bibnamefont{and}
  \bibinfo{author}{\bibfnamefont{M.}~\bibnamefont{Kamionkowski}},
  \bibinfo{journal}{Phys. Rev. D} \textbf{\bibinfo{volume}{94}},
  \bibinfo{pages}{103523} (\bibinfo{year}{2016}), \eprint{1608.01309}.

\bibitem[{\citenamefont{Poulin et~al.}(2019)\citenamefont{Poulin, Smith,
  Karwal, and Kamionkowski}}]{Poulin:2018cxd}
\bibinfo{author}{\bibfnamefont{V.}~\bibnamefont{Poulin}},
  \bibinfo{author}{\bibfnamefont{T.~L.} \bibnamefont{Smith}},
  \bibinfo{author}{\bibfnamefont{T.}~\bibnamefont{Karwal}}, \bibnamefont{and}
  \bibinfo{author}{\bibfnamefont{M.}~\bibnamefont{Kamionkowski}},
  \bibinfo{journal}{Phys. Rev. Lett.} \textbf{\bibinfo{volume}{122}},
  \bibinfo{pages}{221301} (\bibinfo{year}{2019}), \eprint{1811.04083}.

\bibitem[{\citenamefont{Lin et~al.}(2019)\citenamefont{Lin, Benevento, Hu, and
  Raveri}}]{Lin:2019qug}
\bibinfo{author}{\bibfnamefont{M.-X.} \bibnamefont{Lin}},
  \bibinfo{author}{\bibfnamefont{G.}~\bibnamefont{Benevento}},
  \bibinfo{author}{\bibfnamefont{W.}~\bibnamefont{Hu}}, \bibnamefont{and}
  \bibinfo{author}{\bibfnamefont{M.}~\bibnamefont{Raveri}},
  \bibinfo{journal}{Phys. Rev. D} \textbf{\bibinfo{volume}{100}},
  \bibinfo{pages}{063542} (\bibinfo{year}{2019}), \eprint{1905.12618}.

\bibitem[{\citenamefont{Agrawal et~al.}(2019)\citenamefont{Agrawal, Cyr-Racine,
  Pinner, and Randall}}]{Agrawal:2019lmo}
\bibinfo{author}{\bibfnamefont{P.}~\bibnamefont{Agrawal}},
  \bibinfo{author}{\bibfnamefont{F.-Y.} \bibnamefont{Cyr-Racine}},
  \bibinfo{author}{\bibfnamefont{D.}~\bibnamefont{Pinner}}, \bibnamefont{and}
  \bibinfo{author}{\bibfnamefont{L.}~\bibnamefont{Randall}}
  (\bibinfo{year}{2019}), \eprint{1904.01016}.

\bibitem[{\citenamefont{Poulin et~al.}(2018)\citenamefont{Poulin, Smith, Grin,
  Karwal, and Kamionkowski}}]{Poulin:2018dzj}
\bibinfo{author}{\bibfnamefont{V.}~\bibnamefont{Poulin}},
  \bibinfo{author}{\bibfnamefont{T.~L.} \bibnamefont{Smith}},
  \bibinfo{author}{\bibfnamefont{D.}~\bibnamefont{Grin}},
  \bibinfo{author}{\bibfnamefont{T.}~\bibnamefont{Karwal}}, \bibnamefont{and}
  \bibinfo{author}{\bibfnamefont{M.}~\bibnamefont{Kamionkowski}},
  \bibinfo{journal}{Phys. Rev. D} \textbf{\bibinfo{volume}{98}},
  \bibinfo{pages}{083525} (\bibinfo{year}{2018}), \eprint{1806.10608}.

\bibitem[{\citenamefont{Hu et~al.}(2000)\citenamefont{Hu, Barkana, and
  Gruzinov}}]{Hu:2000ke}
\bibinfo{author}{\bibfnamefont{W.}~\bibnamefont{Hu}},
  \bibinfo{author}{\bibfnamefont{R.}~\bibnamefont{Barkana}}, \bibnamefont{and}
  \bibinfo{author}{\bibfnamefont{A.}~\bibnamefont{Gruzinov}},
  \bibinfo{journal}{Phys. Rev. Lett.} \textbf{\bibinfo{volume}{85}},
  \bibinfo{pages}{1158} (\bibinfo{year}{2000}), \eprint{astro-ph/0003365}.

\bibitem[{\citenamefont{Schive et~al.}(2014)\citenamefont{Schive, Chiueh, and
  Broadhurst}}]{Schive:2014dra}
\bibinfo{author}{\bibfnamefont{H.-Y.} \bibnamefont{Schive}},
  \bibinfo{author}{\bibfnamefont{T.}~\bibnamefont{Chiueh}}, \bibnamefont{and}
  \bibinfo{author}{\bibfnamefont{T.}~\bibnamefont{Broadhurst}},
  \bibinfo{journal}{Nature Phys.} \textbf{\bibinfo{volume}{10}},
  \bibinfo{pages}{496} (\bibinfo{year}{2014}), \eprint{1406.6586}.

\bibitem[{\citenamefont{Hui et~al.}(2017)\citenamefont{Hui, Ostriker, Tremaine,
  and Witten}}]{Hui:2016ltb}
\bibinfo{author}{\bibfnamefont{L.}~\bibnamefont{Hui}},
  \bibinfo{author}{\bibfnamefont{J.~P.} \bibnamefont{Ostriker}},
  \bibinfo{author}{\bibfnamefont{S.}~\bibnamefont{Tremaine}}, \bibnamefont{and}
  \bibinfo{author}{\bibfnamefont{E.}~\bibnamefont{Witten}},
  \bibinfo{journal}{Phys. Rev. D} \textbf{\bibinfo{volume}{95}},
  \bibinfo{pages}{043541} (\bibinfo{year}{2017}), \eprint{1610.08297}.

\bibitem[{\citenamefont{Torrado and Lewis}(2020)}]{Torrado:2020dgo}
\bibinfo{author}{\bibfnamefont{J.}~\bibnamefont{Torrado}} \bibnamefont{and}
  \bibinfo{author}{\bibfnamefont{A.}~\bibnamefont{Lewis}}
  (\bibinfo{year}{2020}), \eprint{2005.05290}.

\bibitem[{\citenamefont{Hu et~al.}(1997)\citenamefont{Hu, Sugiyama, and
  Silk}}]{Hu:1995kot}
\bibinfo{author}{\bibfnamefont{W.}~\bibnamefont{Hu}},
  \bibinfo{author}{\bibfnamefont{N.}~\bibnamefont{Sugiyama}}, \bibnamefont{and}
  \bibinfo{author}{\bibfnamefont{J.}~\bibnamefont{Silk}},
  \bibinfo{journal}{Nature} \textbf{\bibinfo{volume}{386}}, \bibinfo{pages}{37}
  (\bibinfo{year}{1997}), \eprint{astro-ph/9604166}.

\bibitem[{\citenamefont{Hu et~al.}(2001)\citenamefont{Hu, Fukugita,
  Zaldarriaga, and Tegmark}}]{Hu:2000ti}
\bibinfo{author}{\bibfnamefont{W.}~\bibnamefont{Hu}},
  \bibinfo{author}{\bibfnamefont{M.}~\bibnamefont{Fukugita}},
  \bibinfo{author}{\bibfnamefont{M.}~\bibnamefont{Zaldarriaga}},
  \bibnamefont{and} \bibinfo{author}{\bibfnamefont{M.}~\bibnamefont{Tegmark}},
  \bibinfo{journal}{Astrophys. J.} \textbf{\bibinfo{volume}{549}},
  \bibinfo{pages}{669} (\bibinfo{year}{2001}), \eprint{astro-ph/0006436}.

\bibitem[{\citenamefont{Hu and Dodelson}(2002)}]{Hu:2001bc}
\bibinfo{author}{\bibfnamefont{W.}~\bibnamefont{Hu}} \bibnamefont{and}
  \bibinfo{author}{\bibfnamefont{S.}~\bibnamefont{Dodelson}},
  \bibinfo{journal}{Ann. Rev. Astron. Astrophys.}
  \textbf{\bibinfo{volume}{40}}, \bibinfo{pages}{171} (\bibinfo{year}{2002}),
  \eprint{astro-ph/0110414}.

\bibitem[{\citenamefont{Aghanim et~al.}(2020{\natexlab{b}})}]{Planck:2018lbu}
\bibinfo{author}{\bibfnamefont{N.}~\bibnamefont{Aghanim}} \bibnamefont{et~al.}
  (\bibinfo{collaboration}{Planck}), \bibinfo{journal}{Astron. Astrophys.}
  \textbf{\bibinfo{volume}{641}}, \bibinfo{pages}{A8}
  (\bibinfo{year}{2020}{\natexlab{b}}), \eprint{1807.06210}.

\bibitem[{\citenamefont{Beutler et~al.}(2011)\citenamefont{Beutler, Blake,
  Colless, Jones, Staveley-Smith, Campbell, Parker, Saunders, and
  Watson}}]{Beutler_2011}
\bibinfo{author}{\bibfnamefont{F.}~\bibnamefont{Beutler}},
  \bibinfo{author}{\bibfnamefont{C.}~\bibnamefont{Blake}},
  \bibinfo{author}{\bibfnamefont{M.}~\bibnamefont{Colless}},
  \bibinfo{author}{\bibfnamefont{D.~H.} \bibnamefont{Jones}},
  \bibinfo{author}{\bibfnamefont{L.}~\bibnamefont{Staveley-Smith}},
  \bibinfo{author}{\bibfnamefont{L.}~\bibnamefont{Campbell}},
  \bibinfo{author}{\bibfnamefont{Q.}~\bibnamefont{Parker}},
  \bibinfo{author}{\bibfnamefont{W.}~\bibnamefont{Saunders}}, \bibnamefont{and}
  \bibinfo{author}{\bibfnamefont{F.}~\bibnamefont{Watson}},
  \bibinfo{journal}{Mon. Not. Roy. Astron. Soc.}
  \textbf{\bibinfo{volume}{416}}, \bibinfo{pages}{3017} (\bibinfo{year}{2011}),
  ISSN \bibinfo{issn}{0035-8711}.

\bibitem[{\citenamefont{Ross et~al.}(2015)\citenamefont{Ross, Samushia,
  Howlett, Percival, Burden, and Manera}}]{Ross:2014qpa}
\bibinfo{author}{\bibfnamefont{A.~J.} \bibnamefont{Ross}},
  \bibinfo{author}{\bibfnamefont{L.}~\bibnamefont{Samushia}},
  \bibinfo{author}{\bibfnamefont{C.}~\bibnamefont{Howlett}},
  \bibinfo{author}{\bibfnamefont{W.~J.} \bibnamefont{Percival}},
  \bibinfo{author}{\bibfnamefont{A.}~\bibnamefont{Burden}}, \bibnamefont{and}
  \bibinfo{author}{\bibfnamefont{M.}~\bibnamefont{Manera}},
  \bibinfo{journal}{Mon. Not. Roy. Astron. Soc.}
  \textbf{\bibinfo{volume}{449}}, \bibinfo{pages}{835} (\bibinfo{year}{2015}),
  \eprint{1409.3242}.

\bibitem[{\citenamefont{Raichoor et~al.}(2020)}]{Raichoor:2020vio}
\bibinfo{author}{\bibfnamefont{A.}~\bibnamefont{Raichoor}}
  \bibnamefont{et~al.}, \bibinfo{journal}{Mon. Not. Roy. Astron. Soc.}
  \textbf{\bibinfo{volume}{500}}, \bibinfo{pages}{3254} (\bibinfo{year}{2020}),
  \eprint{2007.09007}.

\bibitem[{\citenamefont{Bautista et~al.}(2020)}]{Bautista:2020ahg}
\bibinfo{author}{\bibfnamefont{J.~E.} \bibnamefont{Bautista}}
  \bibnamefont{et~al.}, \bibinfo{journal}{Mon. Not. Roy. Astron. Soc.}
  \textbf{\bibinfo{volume}{500}}, \bibinfo{pages}{736} (\bibinfo{year}{2020}),
  \eprint{2007.08993}.

\bibitem[{\citenamefont{Neveux et~al.}(2020)}]{Neveux:2020voa}
\bibinfo{author}{\bibfnamefont{R.}~\bibnamefont{Neveux}} \bibnamefont{et~al.},
  \bibinfo{journal}{Mon. Not. Roy. Astron. Soc.}
  \textbf{\bibinfo{volume}{499}}, \bibinfo{pages}{210} (\bibinfo{year}{2020}),
  \eprint{2007.08999}.

\bibitem[{\citenamefont{du~Mas~des Bourboux
  et~al.}(2020)}]{duMasdesBourboux:2020pck}
\bibinfo{author}{\bibfnamefont{H.}~\bibnamefont{du~Mas~des Bourboux}}
  \bibnamefont{et~al.}, \bibinfo{journal}{Astrophys. J.}
  \textbf{\bibinfo{volume}{901}}, \bibinfo{pages}{153} (\bibinfo{year}{2020}),
  \eprint{2007.08995}.

\bibitem[{\citenamefont{Scolnic et~al.}(2018)}]{Scolnic:2017caz}
\bibinfo{author}{\bibfnamefont{D.~M.} \bibnamefont{Scolnic}}
  \bibnamefont{et~al.}, \bibinfo{journal}{Astrophys. J.}
  \textbf{\bibinfo{volume}{859}}, \bibinfo{pages}{101} (\bibinfo{year}{2018}),
  \eprint{1710.00845}.

\bibitem[{\citenamefont{Riess et~al.}(2019)\citenamefont{Riess, Casertano,
  Yuan, Macri, and Scolnic}}]{Riess:2019cxk}
\bibinfo{author}{\bibfnamefont{A.~G.} \bibnamefont{Riess}},
  \bibinfo{author}{\bibfnamefont{S.}~\bibnamefont{Casertano}},
  \bibinfo{author}{\bibfnamefont{W.}~\bibnamefont{Yuan}},
  \bibinfo{author}{\bibfnamefont{L.~M.} \bibnamefont{Macri}}, \bibnamefont{and}
  \bibinfo{author}{\bibfnamefont{D.}~\bibnamefont{Scolnic}},
  \bibinfo{journal}{Astrophys. J.} \textbf{\bibinfo{volume}{876}},
  \bibinfo{pages}{85} (\bibinfo{year}{2019}), \eprint{1903.07603}.

\bibitem[{\citenamefont{Wong et~al.}(2020)}]{Wong:2019kwg}
\bibinfo{author}{\bibfnamefont{K.~C.} \bibnamefont{Wong}} \bibnamefont{et~al.},
  \bibinfo{journal}{Mon. Not. Roy. Astron. Soc.}
  \textbf{\bibinfo{volume}{498}}, \bibinfo{pages}{1420} (\bibinfo{year}{2020}),
  \eprint{1907.04869}.

\bibitem[{\citenamefont{Reid et~al.}(2009)\citenamefont{Reid, Braatz, Condon,
  Greenhill, Henkel, and Lo}}]{Reid:2008nm}
\bibinfo{author}{\bibfnamefont{M.}~\bibnamefont{Reid}},
  \bibinfo{author}{\bibfnamefont{J.}~\bibnamefont{Braatz}},
  \bibinfo{author}{\bibfnamefont{J.}~\bibnamefont{Condon}},
  \bibinfo{author}{\bibfnamefont{L.}~\bibnamefont{Greenhill}},
  \bibinfo{author}{\bibfnamefont{C.}~\bibnamefont{Henkel}}, \bibnamefont{and}
  \bibinfo{author}{\bibfnamefont{K.}~\bibnamefont{Lo}},
  \bibinfo{journal}{Astrophys. J.} \textbf{\bibinfo{volume}{695}},
  \bibinfo{pages}{287} (\bibinfo{year}{2009}), \eprint{0811.4345}.

\bibitem[{\citenamefont{Freedman et~al.}(2019)}]{Freedman:2019jwv}
\bibinfo{author}{\bibfnamefont{W.~L.} \bibnamefont{Freedman}}
  \bibnamefont{et~al.} (\bibinfo{year}{2019}), \eprint{1907.05922}.

\bibitem[{\citenamefont{Potter et~al.}(2018)\citenamefont{Potter, Jensen,
  Blakeslee et~al.}}]{Potter:2018}
\bibinfo{author}{\bibfnamefont{C.}~\bibnamefont{Potter}},
  \bibinfo{author}{\bibfnamefont{J.~B.} \bibnamefont{Jensen}},
  \bibinfo{author}{\bibfnamefont{J.}~\bibnamefont{Blakeslee}},
  \bibnamefont{et~al.}, \bibinfo{journal}{American Astronomical Society Meeting
  Abstracts \#} \textbf{\bibinfo{volume}{232}}, \bibinfo{pages}{232}
  (\bibinfo{year}{2018}).

\bibitem[{\citenamefont{Huang et~al.}(2018)}]{Huang:2018dbn}
\bibinfo{author}{\bibfnamefont{C.~D.} \bibnamefont{Huang}}
  \bibnamefont{et~al.}, \bibinfo{journal}{Astrophys. J.}
  \textbf{\bibinfo{volume}{857}}, \bibinfo{pages}{67} (\bibinfo{year}{2018}),
  \eprint{1801.02711}.

\bibitem[{\citenamefont{Ade et~al.}(2015)}]{Ade:2014zfo}
\bibinfo{author}{\bibfnamefont{P.~A.~R.} \bibnamefont{Ade}}
  \bibnamefont{et~al.} (\bibinfo{collaboration}{Planck}),
  \bibinfo{journal}{Astron. Astrophys.} \textbf{\bibinfo{volume}{580}},
  \bibinfo{pages}{A22} (\bibinfo{year}{2015}), \eprint{1406.7482}.

\bibitem[{\citenamefont{Hart and Chluba}(2020)}]{Hart:2019dxi}
\bibinfo{author}{\bibfnamefont{L.}~\bibnamefont{Hart}} \bibnamefont{and}
  \bibinfo{author}{\bibfnamefont{J.}~\bibnamefont{Chluba}},
  \bibinfo{journal}{Mon. Not. Roy. Astron. Soc.}
  \textbf{\bibinfo{volume}{493}}, \bibinfo{pages}{3255} (\bibinfo{year}{2020}),
  \eprint{1912.03986}.

\bibitem[{\citenamefont{Hart and Chluba}(2018)}]{Hart:2017ndk}
\bibinfo{author}{\bibfnamefont{L.}~\bibnamefont{Hart}} \bibnamefont{and}
  \bibinfo{author}{\bibfnamefont{J.}~\bibnamefont{Chluba}},
  \bibinfo{journal}{Mon. Not. Roy. Astron. Soc.}
  \textbf{\bibinfo{volume}{474}}, \bibinfo{pages}{1850} (\bibinfo{year}{2018}),
  \eprint{1705.03925}.

\bibitem[{\citenamefont{Riess et~al.}(2021)\citenamefont{Riess, Casertano,
  Yuan, Bowers, Macri, Zinn, and Scolnic}}]{Riess:2020fzl}
\bibinfo{author}{\bibfnamefont{A.~G.} \bibnamefont{Riess}},
  \bibinfo{author}{\bibfnamefont{S.}~\bibnamefont{Casertano}},
  \bibinfo{author}{\bibfnamefont{W.}~\bibnamefont{Yuan}},
  \bibinfo{author}{\bibfnamefont{J.~B.} \bibnamefont{Bowers}},
  \bibinfo{author}{\bibfnamefont{L.}~\bibnamefont{Macri}},
  \bibinfo{author}{\bibfnamefont{J.~C.} \bibnamefont{Zinn}}, \bibnamefont{and}
  \bibinfo{author}{\bibfnamefont{D.}~\bibnamefont{Scolnic}},
  \bibinfo{journal}{Astrophys. J. Lett.} \textbf{\bibinfo{volume}{908}},
  \bibinfo{pages}{L6} (\bibinfo{year}{2021}), \eprint{2012.08534}.

\bibitem[{\citenamefont{Skidmore et~al.}(2015)}]{Skidmore:2015lga}
\bibinfo{author}{\bibfnamefont{W.}~\bibnamefont{Skidmore}} \bibnamefont{et~al.}
  (\bibinfo{collaboration}{TMT International Science Development Teams \& TMT
  Science Advisory Committee}), \bibinfo{journal}{Res. Astron. Astrophys.}
  \textbf{\bibinfo{volume}{15}}, \bibinfo{pages}{1945} (\bibinfo{year}{2015}),
  \eprint{1505.01195}.

\bibitem[{\citenamefont{Behroozi et~al.}(2020)}]{Behroozi:2020jhj}
\bibinfo{author}{\bibfnamefont{P.}~\bibnamefont{Behroozi}}
  \bibnamefont{et~al.}, \bibinfo{journal}{Mon. Not. Roy. Astron. Soc.}
  \textbf{\bibinfo{volume}{499}}, \bibinfo{pages}{5702} (\bibinfo{year}{2020}),
  \eprint{2007.04988}.

\bibitem[{\citenamefont{Lewis et~al.}(2000)\citenamefont{Lewis, Challinor, and
  Lasenby}}]{Lewis:1999bs}
\bibinfo{author}{\bibfnamefont{A.}~\bibnamefont{Lewis}},
  \bibinfo{author}{\bibfnamefont{A.}~\bibnamefont{Challinor}},
  \bibnamefont{and} \bibinfo{author}{\bibfnamefont{A.}~\bibnamefont{Lasenby}},
  \bibinfo{journal}{apj} \textbf{\bibinfo{volume}{538}}, \bibinfo{pages}{473}
  (\bibinfo{year}{2000}), \eprint{astro-ph/9911177}.

\bibitem[{\citenamefont{Hu and Sugiyama}(1996)}]{Hu:1995en}
\bibinfo{author}{\bibfnamefont{W.}~\bibnamefont{Hu}} \bibnamefont{and}
  \bibinfo{author}{\bibfnamefont{N.}~\bibnamefont{Sugiyama}},
  \bibinfo{journal}{Astrophys. J.} \textbf{\bibinfo{volume}{471}},
  \bibinfo{pages}{542} (\bibinfo{year}{1996}), \eprint{astro-ph/9510117}.

\bibitem[{\citenamefont{Gil-Mar\'\i{}n et~al.}(2016)}]{Gil-Marin:2015nqa}
\bibinfo{author}{\bibfnamefont{H.}~\bibnamefont{Gil-Mar\'\i{}n}}
  \bibnamefont{et~al.}, \bibinfo{journal}{Mon. Not. Roy. Astron. Soc.}
  \textbf{\bibinfo{volume}{460}}, \bibinfo{pages}{4210} (\bibinfo{year}{2016}),
  \eprint{1509.06373}.

\bibitem[{\citenamefont{Kessler and Scolnic}(2017)}]{Kessler:2016uwi}
\bibinfo{author}{\bibfnamefont{R.}~\bibnamefont{Kessler}} \bibnamefont{and}
  \bibinfo{author}{\bibfnamefont{D.}~\bibnamefont{Scolnic}},
  \bibinfo{journal}{Astrophys. J.} \textbf{\bibinfo{volume}{836}},
  \bibinfo{pages}{56} (\bibinfo{year}{2017}), \eprint{1610.04677}.

\bibitem[{\citenamefont{Dainotti et~al.}(2021)\citenamefont{Dainotti,
  De~Simone, Schiavone, Montani, Rinaldi, and Lambiase}}]{Dainotti:2021pqg}
\bibinfo{author}{\bibfnamefont{M.~G.} \bibnamefont{Dainotti}},
  \bibinfo{author}{\bibfnamefont{B.}~\bibnamefont{De~Simone}},
  \bibinfo{author}{\bibfnamefont{T.}~\bibnamefont{Schiavone}},
  \bibinfo{author}{\bibfnamefont{G.}~\bibnamefont{Montani}},
  \bibinfo{author}{\bibfnamefont{E.}~\bibnamefont{Rinaldi}}, \bibnamefont{and}
  \bibinfo{author}{\bibfnamefont{G.}~\bibnamefont{Lambiase}},
  \bibinfo{journal}{Astrophys. J.} \textbf{\bibinfo{volume}{912}},
  \bibinfo{pages}{150} (\bibinfo{year}{2021}), \eprint{2103.02117}.

\bibitem[{\citenamefont{Chluba and Thomas}(2011)}]{Chluba:2010ca}
\bibinfo{author}{\bibfnamefont{J.}~\bibnamefont{Chluba}} \bibnamefont{and}
  \bibinfo{author}{\bibfnamefont{R.}~\bibnamefont{Thomas}},
  \bibinfo{journal}{Mon. Not. Roy. Astron. Soc.}
  \textbf{\bibinfo{volume}{412}}, \bibinfo{pages}{748} (\bibinfo{year}{2011}),
  \eprint{1010.3631}.

\bibitem[{\citenamefont{Ali-Haïmoud and Hirata}(2011)}]{Ali_Ha_moud_2011}
\bibinfo{author}{\bibfnamefont{Y.}~\bibnamefont{Ali-Haïmoud}}
  \bibnamefont{and} \bibinfo{author}{\bibfnamefont{C.~M.}
  \bibnamefont{Hirata}}, \bibinfo{journal}{Physical Review D}
  \textbf{\bibinfo{volume}{83}} (\bibinfo{year}{2011}), ISSN
  \bibinfo{issn}{1550-2368},
  \urlprefix\url{http://dx.doi.org/10.1103/PhysRevD.83.043513}.

\bibitem[{\citenamefont{Lee and Ali-Ha\"\i{}moud}(2020)}]{Lee:2020obi}
\bibinfo{author}{\bibfnamefont{N.}~\bibnamefont{Lee}} \bibnamefont{and}
  \bibinfo{author}{\bibfnamefont{Y.}~\bibnamefont{Ali-Ha\"\i{}moud}},
  \bibinfo{journal}{Phys. Rev. D} \textbf{\bibinfo{volume}{102}},
  \bibinfo{pages}{083517} (\bibinfo{year}{2020}), \eprint{2007.14114}.

\bibitem[{\citenamefont{Blas et~al.}(2011)\citenamefont{Blas, Lesgourgues, and
  Tram}}]{Blas_2011}
\bibinfo{author}{\bibfnamefont{D.}~\bibnamefont{Blas}},
  \bibinfo{author}{\bibfnamefont{J.}~\bibnamefont{Lesgourgues}},
  \bibnamefont{and} \bibinfo{author}{\bibfnamefont{T.}~\bibnamefont{Tram}},
  \bibinfo{journal}{Journal of Cosmology and Astroparticle Physics}
  \textbf{\bibinfo{volume}{2011}}, \bibinfo{pages}{034–034}
  (\bibinfo{year}{2011}), ISSN \bibinfo{issn}{1475-7516},
  \urlprefix\url{http://dx.doi.org/10.1088/1475-7516/2011/07/034}.

\bibitem[{\citenamefont{Luu}(2021)}]{Luu:2021yhl}
\bibinfo{author}{\bibfnamefont{H.~N.} \bibnamefont{Luu}}
  (\bibinfo{year}{2021}), \eprint{2111.01347}.

\bibitem[{\citenamefont{Pogosian et~al.}(2020)\citenamefont{Pogosian, Zhao, and
  Jedamzik}}]{Pogosian:2020ded}
\bibinfo{author}{\bibfnamefont{L.}~\bibnamefont{Pogosian}},
  \bibinfo{author}{\bibfnamefont{G.-B.} \bibnamefont{Zhao}}, \bibnamefont{and}
  \bibinfo{author}{\bibfnamefont{K.}~\bibnamefont{Jedamzik}},
  \bibinfo{journal}{Astrophys. J. Lett.} \textbf{\bibinfo{volume}{904}},
  \bibinfo{pages}{L17} (\bibinfo{year}{2020}), \eprint{2009.08455}.

\end{thebibliography}

\newoutputstream{stream}
\openoutputfile{counters3}{stream}
\addtostream{stream}{
  \protect\setcounter{equation}{\arabic{equation}}}
\addtostream{stream}{
  \protect\setcounter{table}{\arabic{table}}}  
\addtostream{stream}{
  \protect\setcounter{figure}{\arabic{figure}}}    
\closeoutputstream{stream}

\end{document}